\documentclass{article}
\usepackage[T1]{fontenc}
\usepackage{graphicx} 
\usepackage[margin=1in]{geometry}
\usepackage{setspace}
\usepackage[authoryear]{natbib}
\usepackage[hidelinks,colorlinks=true,linkcolor=blue,citecolor=blue]{hyperref}
\usepackage{amssymb, amsmath}
\usepackage{booktabs}
\usepackage{float}

\renewcommand{\hat}{\widehat}
\renewcommand{\tilde}{\widetilde}

\makeatletter 
\newenvironment{keyword}{
  \par\noindent\textbf{Keywords: } 
}{
  \par 
} 
\newcommand{\kwd}[1]{#1} 
\makeatother

\makeatletter \@ifundefined{acks}{ 
\newenvironment{acks}[1][Acknowledgments]{ 
  \section*{#1}
   }{ 
   } 
}{ 
  \renewenvironment{acks}[1][Acknowledgments]{ 
  \section*{#1} 
  }{ 
  } 
} 
\makeatother

\title{Nonprobability Samples for Small Area Estimation: A Review and Comparative Simulation Study}
\author{ Sho Kawano\thanks{University of California, Santa Cruz}
\and Daniel Vedensky\thanks{Simon Fraser University. Corresponding author: \url{daniel_vedensky@sfu.ca}}
\and Qianyu Dong\footnotemark[1] 
\and Ethan Pawl\footnotemark[1] 
\and Qi Wang\thanks{Emory University} 
\and Paul A. Parker\footnotemark[1] 
\and Zehang Richard Li\footnotemark[1] 
\and Scott H. Holan\thanks{University of Missouri}}

\begin{document}

\maketitle
\makeatletter \begingroup 
\renewcommand\@makefnmark{} 
\footnotetext{Sho Kawano and Daniel Vedensky contributed equally to this manuscript.} 
\endgroup 
\makeatother
 \abstract{ 
 Nonprobability samples (NPS) are attractive because they are less costly to collect, can provide substantially larger sample sizes, and may reach populations that traditional probability surveys do not.
As response rates for traditional surveys fall, interest in NPS has grown rapidly within the field of survey statistics.
These methods are especially relevant for small area estimation (SAE), where there is ever-present demand for estimates at fine geographic scales and detailed demographic domains.
Despite rapid methodological development, there remains limited understanding of which approaches perform best under different conditions.
In this paper, we review recent developments in NPS methodology, including the concept of data defect correlation (DDC) as a measure of data quality and as a tool for categorizing the various NPS methods.
We then present a comprehensive simulation study that evaluates a range of NPS approaches under varying levels of DDC and extend several existing methods to the SAE setting.

 \\
 }

\doublespacing
\section{Introduction}

Declining response rates and changes in technology have caused a dramatic rise in the cost of collecting probability sample (PS) data \citep{beaumont_2020}.
These fundamental pressures on traditional survey data collection have driven a renewed interest in the use of nonprobability sample (NPS) data for official statistics.
This renewed interest is especially relevant for small area estimation (SAE), where there is ever-present demand to produce estimates at finer geographies and for more detailed demographic domains.
NPS data offer a potentially valuable source of information for SAE problems, yet the connection between the two areas remains comparatively underexplored.
To help bridge this gap, this paper reviews connections that have been made in the literature, proposes extensions of existing NPS methods to SAE, and presents a comprehensive simulation study to evaluate when different methods are most appropriate.

NPS refers to any sample drawn without known inclusion probabilities.
Data collected in this fashion differ from the traditional PS data typically used in official statistics, where people or households are sampled according to a known design.
Most notably, NPS methods include online opt-in web panels, but other common ``designs'' include quota sampling and network sampling \citep{elliott_2017}.
NPS offers a promising avenue to fielding surveys at reduced cost, with shorter turnaround time, and with much larger sample sizes \citep{fulop_2022}.
In many cases, NPS methods may now be the only way to reach large segments of the population \citep{bailey_2022}.
In addition, NPS methods have found use in areas that PS methods are not well-suited for, such as measuring hard-to-reach populations \citep{jerit_2023, klinke_2025} or surveying people in non-democratic countries \citep{korsunava_2025}.
These advantages have led to a proliferation of research into NPS methods, including a raft of recent review papers \citep{elliott_2017, zhang2019, wu2022} and software for conducting NPS analyses \citep{nonprobest, nonprobsvy, nonprobsampling}.

Two major developments in the analysis of NPS data are central to our framing.
First, \cite{meng2018} introduced the data defect correlation (DDC) as a measure of data quality.
Second, \cite{wu2022} applied this concept to survey methodology showing that NPS methods can generally be classified into three categories: inverse probability weighting (IPW), superpopulation (SP), and doubly robust (DR).
Each category can be interpreted as reducing DDC, and thereby improving estimates, in a specific way.
This line of research has begun to clarify under what conditions NPS methods are able to perform well \citep{zhang2019, meng2022} and when they are ``fit for purpose'' \citep{jerit_2023}.
We review these concepts in Section~\ref{sec:overview} and make DDC a central part of the simulation study presented in Section~\ref{sec:sim}.
The simulation study also examines several other important distinctions in classifying NPS methods.
One is whether the response variable is available in both the PS and NPS or only the NPS.
Another is whether the method requires population-level information such as unit-level covariates or population totals.
While many NPS survey methods assume a reference PS is available to aid estimation, we also compare an alternative that relies only on NPS data and population auxiliary information.

While there is a long history of SAE methods incorporating alternative datasets, NPS data strategies are different. 
For instance, early SAE methods made use of satellite imagery \citep{battese1988} and more recent work has used web data \citep{porter_2014} and mobility data \citep{marchetti_2015}.
These strategies generally use the alternative data to construct covariates only and this differs from most of the DDC-minimizing techniques we examine. 
Nevertheless, bringing developments in NPS methodology to the SAE sphere can be viewed as a natural next step in the field's well-established use of big data and one that is crucial to investigate.

Recent SAE work has begun to make this connection between the DDC framework and SAE explicit \citep{rao_2020}.
Some recent works even couch their contribution explicitly in terms of DDC, such as the DR method proposed by \cite{schirripa_spagnolo2024} and the MRP extensions proposed by \cite{si_2025}.
However, not all of the DDC-minimizing categories have been explored in the context of SAE.
Meanwhile, newer methods that do not fit into these categories have appeared.
For example, \cite{aliste2025combining} model a bias term rather than a survey response, while \cite{salvatore_2023} incorporate NPS data into a Bayesian prior. 
Given the rapid pace of development,  many of these methods have not been directly compared against each other and there is a need to develop guidelines for when to employ certain methods over others.

In order to provide such guidance and assess which methods are most appropriate in which settings, this paper presents a review and empirical simulation study of NPS SAE methods.
Using data from the American Community Survey (ACS), we construct a realistic finite population from which we draw both probability and nonprobability samples.
By varying the value of the DDC in the NPS sample, we examine how data quality affects the usefulness of NPS data and affects recommendations of which methods to use.
We evaluate the performance of representative models from each category and propose several extensions including an extension of the NPS-informed prior method of \cite{salvatore_2023} to the SAE setting as well as a Bayesian, unit-level IPW model.

The rest of the paper proceeds as follows.
In Section~\ref{sec:overview} we introduce the notation and necessary technical background, including the DDC framework of \cite{meng2018}.
We then describe each of the major NPS categories and our modifications.
Section~\ref{sec:sim} presents the empirical simulation study using ACS PUMS (Public Use Microdata Sample) data and interprets the results.
Section~\ref{sec:discussion} concludes with a discussion.

\section{Overview of NPS methods} \label{sec:overview}
\subsection{Notation}

Let $U$ be a finite population of size $N$ with units indexed by $i=1,\dots,N$ and partitioned into domains (or areas) $d=1,\dots,D$.
We write $U_d$ for the set of units in domain~$d$ and $N_d = |U_d|$ for the size of the domain so that $\bigcup_{d=1}^D U_d = U$ and $N=\sum_{d=1}^D N_d$.
When domain-level quantities are of interest (e.g., small-area means), all definitions below apply within a domain by restricting indices to units in that domain.

Let $y_i$ denote the outcome of interest for unit~$i$, with finite-population mean $\bar y$. 
For simplicity, we focus on binary outcomes, which are more common in the survey setting. 
However, most methods under consideration apply to totals and means of other data types.
We note, too, that most existing NPS methods apply to means, but some work has begun to consider estimating general parameters \citep{chen_haziza_2022}, including quantiles \citep{berg_2025}.

Each unit is associated with a $q-$dimensional covariate vector $\mathbf{x}_i$ and a domain assignment $d\in\{1,\ldots, D\}$.
We use bracket notation to pick out a unit's domain, so that $d[i]$ equals the index of the domain to which unit $i$ belongs.

Furthermore, let $r_i\in\{0,1\}$ indicate whether unit $i$ is included in the probability sample, so that the realized sample set is $S=\{i:r_i=1\}$ with sample size $n = |S| = \sum_{i=1}^N r_i,$ 
and sampling fraction $f = \frac{n}{N}.$
 We denote the sample size in domain $d$ by $n_d.$
Each sampled unit carries a known design weight $w_i > 0$, typically the inverse of its inclusion probability $\pi_i = \Pr(r_i = 1)$.

We also define the analogous quantities for an NPS. 
Let $r_i^*\in\{0,1\}$ indicate whether unit $i$ is included in the nonprobability sample, with observed sample, sample size, and sampling fraction
\[ S^*=\{i:r_i^*=1\}, \quad n^* = |S^*| = \sum_{i=1}^N r_i^*, \quad f^* = \frac{n^*}{N},\]
respectively.

In contrast to the PS, the NPS inclusion mechanism is unknown to the analyst.
We write $\pi_i^* = \Pr(r_i^* = 1 \mid \mathbf{x}_i)$ for the (unknown) propensity of inclusion in the NPS conditional on covariates, following \cite{elliott_2017}.
An asterisk superscript marks NPS-specific quantities (i.e., $r_i^*, S^*, \pi_i^*$, $n^*$, $f^*$, and $w^*_i$) to distinguish them from their PS counterparts.

The covariates $\mathbf{x}_i$ are assumed to be observed for all units in both samples $S$ and $S^*$.
The outcome $y_i$ is always observed in the NPS ($i \in S^*$), but not necessarily in the PS. 
Certain methods, notably superpopulation and doubly robust estimators (Sections~\ref{sec:sp}--\ref{sec:dr}), require population-level covariate information, either in the form of a complete frame $\{\mathbf{x}_i\}_{i=1}^N$ or as marginal or cross-classified cell counts.
We state the specific data requirements for each method as it is introduced.

\subsection{Basic Estimators}
The unweighted sample mean for the PS is
\[
\hat{\bar y}_r
= \frac{1}{n}\sum_{i\in S} y_i = \frac{\sum_{i=1}^N r_i y_i}{\sum_{i=1}^N r_i}
\]
and the analogous quantity for the NPS
\[\hat{\bar y}_{r^*} = \frac{1}{n^*}\sum_{i\in S^*} y_i,\]
is defined identically with $r_i^*$ replacing $r_i$.
Using the PS design weights, a weighted estimator of the finite-population mean is the Horvitz--Thompson estimator \citep{HorvitzThompson}.
\[
\hat{\bar y}_{\mathrm{HT}} = \frac{1}{N}\sum_{i\in S} w_i y_i.
\]
Alternatively, replacing $N$ by the estimated population size $\hat N_w = \sum_{i\in S} w_i$ yields the H\'ajek estimator \citep{hajek_1971}
\[
\hat{\bar y}_{\mathrm{H}} = \frac{1}{\hat{N}_w}\sum_{i\in S} w_i y_i.
\]
If the weights are scaled (or calibrated) so that $\hat N_w = N$, then $\hat{\bar y}_{\mathrm{HT}}=\hat{\bar y}_{\mathrm{H}}$.
Since the NPS has no known design weights, these weighted estimators are not directly available for the NPS. 
Constructing appropriate weights for the NPS is the subject of the IPW methods reviewed in Section~\ref{sec:ipw}.

\subsection{Data Defect Correlation and a Bias-aware Effective Sample Size}\label{sec:ddc}
\cite{meng2018, meng2022} introduced the \textit{data defect correlation} (DDC), defined as the finite-population correlation between the inclusion indicators and the outcome of interest for a given sample
\[\rho_{r,y} = \mathrm{Corr}_U(r,y).\]
The identity holds for any sample, whether obtained through PS or NPS, though we can distinguish the two by referring to
$\rho_{r^*, y^*} = \mathrm{Corr}_U(r^*,y^*)$ for the NPS.
The DDC appears in an exact finite-population expression for the error of the unweighted sample mean \citep{meng2018}
\begin{equation}
\label{eq:ddc_identity}
\hat{\bar y}_{r^*} - \bar y
=
\rho_{r^*\!,y}\,
\sqrt{\frac{1-f^*}{f^*}}\,
\sigma_y,
\end{equation}
where $\sigma_y=\sqrt{\mathrm{Var}_{U}(y)}$ and $\mathrm{Var}_{U}(r^*)=f^*(1-f^*)$.
The three components are (i) \emph{data quality} through $\rho_{r^*\!,y}$,
(ii) \emph{data quantity} through $\sqrt{(1-f^*)/f^*}$, and (iii) \emph{problem difficulty} through $\sigma_y$.
The data quality is the only one of these components that an analyst has control over.
Yet, $\rho_{r^*\!,y}$ cannot be identified from the NPS alone without additional information or modeling assumptions.

For a probability sample, $\rho_{r,y}$ is $O_p(N^{-1/2})$ by design \citep[Theorem~1]{meng2018}. 
This is the formal sense in which a probability sampling design itself reduces DDC. 
An NPS has no such guarantee and the methods we review below confront the resulting bias in various ways.

\cite{meng2018} also presents a \emph{bias-aware} effective sample size $n_{\mathrm{eff}}$ that is defined by matching the mean squared error (MSE) produced by taking the expectation of identity (\ref{eq:ddc_identity}) to that of a perfectly representative simple random sample.
Ignoring finite-population corrections, this yields the approximation
\[
n^*_{\mathrm{eff}}
\approx
\frac{f^*}{1-f^*}\,\frac{1}{\rho_{r^*\!,y}^2}.
\]
This approximation differs from the usual effective sample size \citep{kish_1965}, which reflects variance inflation from weight variability but does not address residual selection bias (i.e., it does not incorporate $\rho_{r^*\!,y}$).
The formula highlights that even a small DDC can severely reduce $n^*_{\mathrm{eff}}$ when $N$ is large.
For example, taking $\rho_{r^*\!,y}=0.005$ and $f^*=0.01$ (suppose $N=300,000,000$ and $n^*=3,000,000$) we obtain roughly $n^*_{\mathrm{eff}} = 400$.
\cite{meng2018} uses this phenomenon to illustrate ``a \textit{Big Data Paradox}: the more the data, the surer we fool ourselves."

\subsection{DDC Reduction}\label{sec:ddc-strategies}

Given the severity of even small DDC values demonstrated above, an important question is how methods that combine PS and NPS data can address NPS bias.
Recall that probability sampling achieves $\rho_{r,y} = O_p(N^{-1/2})$ by design. \cite{meng2022} defines \emph{miniaturizing} the DDC as bringing $|\rho_{r^*\!,y}|$ down to this order for a nonprobability sample.
\cite{meng2022}, identifies two strategies available for doing so.
Identity (\ref{eq:ddc_identity}) expresses the error of the \emph{unweighted} NPS mean through the correlation $\rho_{r^*\!,y}$.
For a \emph{weighted} mean the analogous expression involves the finite-population covariance $\mathrm{Cov}_U$ directly.
Consider a generic weighted mean of NPS observations written in scaled form,
\[
\hat{\bar y}_{\tilde w}
=
\frac{1}{N}\sum_{i=1}^N r_i^* \tilde w_i y_i,
\qquad\text{with}\qquad
\frac{1}{N}\sum_{i=1}^N r_i^* \tilde w_i = 1,
\]
where $\tilde w_i$ denotes an arbitrary weight for NPS unit $i$ (not to be confused with the PS design weight $w_i$ of Section~2.1).
Then, $\hat{\bar y}_{\tilde w}-\bar y = \mathrm{Cov}_{U}(r^* \tilde w, y)$.
To model NPS inclusion, recall $\pi_i^* = \Pr(r_i^* = 1 \mid \mathbf{x}_i)$ from Section~2.1. Under a quasi-randomization model (i.e., assuming that, conditional on $\mathbf{x}_i$, units are included independently with $\mathbb{E}(r_i^*\mid \mathbf{x}_i)=\pi_i^*$) the model-based bias is governed by $\mathrm{Cov}_{U}(\pi^* \tilde w, y)$.
Let $m(\cdot)$ denote a working outcome regression function and let the residuals be
$e_i = y_i - m(\mathbf{x}_i)$.
Writing $m(\mathbf{x})$ as shorthand for the finite-population
sequence $\{m(\mathbf{x}_i)\}_{i=1}^N$, we have
\begin{equation}
\label{eq:cov_decomposition}
\mathrm{Cov}_{U}(\pi^* \tilde w, y)
=
\mathrm{Cov}_{U}(\pi^* \tilde w, m(\mathbf{x}))
+
\mathrm{Cov}_{U}(\pi^* \tilde w, e).
\end{equation}
This decomposition identifies the two strategies available for reducing DDC:
\begin{enumerate}
    \item Choose $\tilde w_i$ so that $\pi_i^* \tilde w_i \propto 1$, which shrinks both covariance terms.
    This is the logic behind IPW and related calibration ideas (Section~\ref{sec:ipw}).
    \item  Choose $m(\mathbf{x}_i)\approx \mathbb{E}(y_i\mid \mathbf{x}_i)$ so that residuals $e_i$
    carry little remaining selection association.
    This is the logic behind superpopulation methods, including mass imputation (Section~\ref{sec:sp-mi}) and MRP (Section~\ref{sec:sp-mrp}).
\end{enumerate}
Doubly robust (DR) estimators combine both strategies: if either $\pi_i^* \tilde w_i$ is approximately constant
or $\mathrm{Cov}_{U}(\pi^* \tilde w, e)$ is approximately zero, then the remaining outcome-relevant selection can be small. 
Some MRP methods also combine both strategies, though they do not formally meet the doubly robust criteria discussed in Section~\ref{sec:dr}.

\subsection{Inverse Probability Weighting (IPW)}\label{sec:ipw}
IPW methods, often also referred to as quasirandomization methods, estimate inclusion probabilities for the NPS so that it can be analyzed using PS techniques that require design weights.
A model for the NPS inclusion probability $\pi_i^*$ is fit using information from the PS, and the resulting weights $\hat{w}_i^* = 1/\hat{\pi}_i^*$ are applied to NPS observations. 
When both samples record the outcome variable, the weighted NPS can be combined with the PS for joint estimation of finite population quantities.
These methods are appealing for their simplicity and the fact that, once weights are estimated, many familiar PS methods apply.

Several methods exist for estimating $\pi_i^*$ \citep{elliott_2017, beresovsky_2026}.
One common set of methods begins by combining both samples and defining a membership indicator $Z_i = 1$ for $i \in S^*$ and $Z_i = 0$ for $i \in S$.
If the PS design is ignorable and a common set of covariates is known for both samples, $\pi_i^*$ can be estimated via unweighted logistic regression of $Z_i$ on $\mathbf{x}_i$.
The fitted probabilities $\hat{P}(Z_i = 1 \mid \mathbf{x}_i)$ estimate NPS inclusion conditional on membership in the combined sample. 
The NPS weights are then taken to be $\hat{w}_i^* = \hat{P}(Z_i = 0 \mid \mathbf{x}_i) / \hat{P}(Z_i = 1 \mid \mathbf{x}_i)$.

When the PS design is not ignorable, the PS design weights are incorporated to estimate $P(r_i^* = 1 \mid \mathbf{x}_i)$ unconditionally (that is, not conditioned on membership in the combined sample).
In this case, a logistic regression of $Z_i$ on $\mathbf{x}_i$ is fit with weights
\begin{align}
\label{eq:ipw_weights}   
\omega_i = \begin{cases}
w_i &\text{ for } i\in S\\
1 &\text{ for } i\in S^*
\end{cases}
\end{align}
and the estimated propensities are used to obtain $\hat{w}_i^* = 1/\hat{\pi}_i^*$.

As an alternative to modeling the sample membership indicator directly, \cite{elliott_2017} propose regressing the PS inclusion probabilities $\hat{\pi}_i = 1/w_i$ on $\mathbf{x}_i$ via beta regression.
The fitted model is then used to predict $\hat{\pi}_i^*$ for $i \in S^*$, and weights are constructed as $\hat{w}_i^* = 1/\hat{\pi}_i^*$.

Most IPW procedures build off of the above approach \citep{wang_2021}.
\cite{chen_et_al2019} propose a more robust IPW estimator that employs a pseudolikelihood to estimate inclusion probabilities and has an analytic expression for the variance.
Namely, they assume $\pi_i = \text{logit}^{-1}(\mathbf{x}_i^\top\pmb{\beta})$ and solve the score equation
\[\frac{1}{N}\left(\sum_{i\in S^*}\mathbf{x}_i - \sum_{i\in S}w_i \pi_i\mathbf{x}_i\right) = 0\]
for $\pmb{\beta}$ then set $\hat{w}^*_i =1+1/\exp^{-1}(\mathbf{x}_i^{\top}\pmb{\beta})$ for $i\in S^*.$
If $\pi_i$ can be assumed to be small, \cite{wang_2021} propose modeling transformed propensity scores $\tilde{\pi}_i = \pi_i/(1+\pi_i)$
and instead solve the score equation
\[\frac{1}{N+n^*}\left(\sum_{i\in S^*}(1-\tilde{\pi}_i)\mathbf{x}_i - \sum_{i\in S}w_i\tilde{\pi}_i\mathbf{x}_i\right)=0.\]
They also point out rescaling the original survey weights can increase efficiency. 
\cite{elliott_2017, chen_2019, wang_2021} all assume the overlap in PS and NPS is negligible, but taking the overlap into account can improve estimates \citep{beresovsky_2026}.

If only the NPS is to be analyzed, the estimated weights can be used directly.
When the response is available in both samples, the PS and NPS weights are combined for a unified analysis.
However, because the NPS is typically much larger, the combined weights must be rescaled to balance the contributions of the two samples.

The resulting weights can be incorporated into an HT or H\'ajek direct estimator.
In the SAE setting, such an estimator is likely to be unstable as some domains may have low sample size.
It is then helpful to model additional structure such as a domain-level random effect.
To do so, the weights can be included in a Fay-Herriot area-level model \citep{ciginas_2026} or a unit-level pseudo-likelihood model \citep{parker2022}.
In the case where many domains have no sampled units, the latter is often a more practical choice.

For a binary outcome, a Bayesian pseudo-likelihood unit-level model takes the form of a binomial likelihood with a logit link
\begin{align}
\label{eq:basic_unit_level}
\tilde{p}(\mathbf{y} \mid \boldsymbol{\beta}, \boldsymbol{\eta}) &\propto \prod_{i\in S} \bigl[p_i^{y_i}(1-p_i)^{1-y_i}\bigr]^{w_i}\\
& \quad\times \prod_{i\in S^*} \bigl[p_i^{y_i}(1-p_i)^{1-y_i}\bigr]^{\hat{w}_i^*},\nonumber \\
\mathrm{logit}(p_i) &= \mathbf{x}_i^\top\boldsymbol{\beta} + \eta_{d[i]},\nonumber
\end{align}
where $\boldsymbol{\beta}$ is a vector of fixed-effect coefficients and $\boldsymbol{\eta} = (\eta_1,\dots,\eta_D)^\top$ is a vector of domain-level random effects.
Section~\ref{sec:sim-methods} describes the weight rescaling and prior specifications.
Model parameters can be estimated with Markov chain Monte Carlo (MCMC) and population quantities are calculated by combining the parameter estimates with population covariate information.
Specifically, for a binary model and MCMC parameter estimates $\hat{\mathbf{\beta}}^{(t)}$ and $\hat{\mathbf{\eta}}^{(t)}$, we sample population values 
\[\tilde{y}_i^{(t)} \sim \text{Bernoulli}\left(\text{logit}^{-1}(\mathbf{x}_i^\top\hat{\boldsymbol{\beta}}^{(t)} + \hat{\boldsymbol{\eta}}_{d[i]}^{(t)})\right)\]
for $i=1,\ldots, N_d.$
This generated population can then be grouped by domain and population quantities within each domain calculated.   
Doing so for each MCMC iteration $t=1,\ldots, T$ allows for the calculation of posterior means and credible intervals of domain estimates. 

\subsection{Superpopulation Methods (SP)}\label{sec:sp}

SP methods specify an outcome model $m(\mathbf{x}_i)\approx \mathbb{E}(y_i\mid \mathbf{x}_i)$ and use predictions from the model to reconstruct population quantities \citep{elliott_2017, zhang2019, wu2022}.
The key assumption is one of \textit{transportability}: that the conditional distribution of $y_i$ given $\mathbf{x}_i$, as estimated from the NPS, is the same as in the target population \citep{kim2021}.
If $\hat m$ is well-specified, the residuals $e_i = y_i - m(\mathbf{x}_i)$ carry little association with the NPS inclusion indicator $r_i^*$, so that $\mathrm{Cov}_{U}(\pi^* \tilde w, e)$ in the decomposition (\ref{eq:cov_decomposition}) is small.

The primary way in which SP methods differ from each other lies in how they specify $m$ and how they perform population-level aggregation.
We review two approaches below.
Mass imputation is not included in our simulation study but illustrates the prediction-and-aggregation logic common to all SP methods, providing context for MRP.
MRP extends this structure with hierarchical shrinkage and is included in the simulation.

\subsubsection{Mass Imputation}\label{sec:sp-mi}

\cite{kim2021} propose a mass imputation estimator that fits an outcome model on the NPS, imputes the outcome for PS units (where the response is assumed not to be observed), and then estimates the population mean using design-weighted averages of the imputed values.
Concretely, let $\hat m(\mathbf{x}_i)$ denote a prediction from a model fit to $(y_i, \mathbf{x}_i)$ for $i \in S^*$.
The mass imputation estimator of the finite-population mean is
\[
\hat{\bar y}_{\mathrm{MI}}
=
\frac{1}{N}\sum_{i\in S} w_i \,\hat m(\mathbf{x}_i),
\]
or the H\'ajek variant $\hat{\bar y}_{\mathrm{MI,H}} = \sum_{i\in S} w_i \,\hat m(\mathbf{x}_i) \big/ \sum_{i\in S} w_i$ when $N$ is unknown.
This estimator requires $\mathbf{x}_i$ to be observed in both samples but does not assume that $y_i$ is observed in the PS.

The consistency of $\hat{\bar y}_{\mathrm{MI}}$ depends on the transportability assumption. 
If this holds and $\hat m$ is consistent for $m$, then the design-weighted average of $\hat m(\mathbf{x}_i)$ over the PS is a consistent estimator of $\bar y$ since the PS is itself representative by design.
The outcome model absorbs the dependence between $y_i$ and $r_i^*$ (the second strategy of Section~\ref{sec:ddc-strategies}), while the PS weights handle the remaining aggregation to the population.

\subsubsection{Multilevel Regression and Poststratification (MRP)}\label{sec:mrp}\label{sec:sp-mrp}

MRP \citep{gelman1997poststratification} extends the mass imputation framework by partitioning the population into $J$ poststratification cells defined by the cross-classification of discrete covariates, then fitting a multilevel regression model to the outcome and aggregating predictions over the cells rather than over PS units.
The MRP estimator for binary data obtained from a PS \citep{park_gelman_bafumi_2004} for a domain $d$ takes the form 
\begin{align}   
\label{eq:mrp_poststratification}
\hat{\bar{y}}_d = \frac{1}{N_d}\sum_{j\in \mathcal{J}_d} {N_j}\hat{p}_j,
\end{align}
where $\mathcal{J}_d$ is the set of poststratification cells belonging to domain $d$, $N_j$ is the population count in cell $j$, and $\hat{p}_j$ is the proportion estimated by the multi-level regression model.
The multilevel regression model introduces partial pooling across cells \citep{gelman_hill_2006, si_2025}.

In the context of NPS data, \cite{si_2025} presents a MRP data integration framework, referred to as MRP-INT-P, that explicitly models NPS inclusion probabilities and incorporates them as predictors in the outcome model. 
\cite{si_2025} estimates the inclusion probabilities using only the NPS.
This inclusion model assumes the cell NPS sample sizes are distributed multinomial
\begin{align*}
(&n^*_1,\ldots, n^*_J\mid \pi^*_1,\ldots, \pi^*_J, N_1, \ldots, N_J)\\
 &\sim \text{Mult}\left(\left(\frac{N_1\pi^*_1}{\sum_{j=1}^J N_j\pi^*_j},\ldots, \frac{N_J\pi^*_J}{\sum_{j=1}^J N_j\pi^*_j}\right), n^*\right) 
\end{align*}
and the inclusion probabilities are modeled as a function of cell-level covariates, $\boldsymbol{\gamma}_j$ 
\begin{align}
\label{eq:mrp_inclusion_model}
\pi^*_j = \text{logit}^{-1}(\boldsymbol{\gamma}_j^\top\alpha).
\end{align} 
Because the NPS is typically large, inclusion probabilities estimated in this way can be more flexible than the IPW methods of Section~\ref{sec:ipw}, which rely on the smaller PS (e.g., a PS may not have enough data to estimate an interaction term).

The estimated NPS inclusion probabilities $\hat{\pi}^*_j$, are then used as predictors in the MRP outcome model 
\begin{align*}
(y_{i} \mid \hat{\pi}^*_{j[i]}, \mathbf{x}_{j[i]}, \boldsymbol{\beta}) &\sim \text{Bernoulli}(p_{i})\\
\text{logit}(p_{i}) &= f(\hat{\pi}^*_{j[i]}) + \mathbf{x}_{j[i]}^\top\boldsymbol{\beta}
\end{align*}
where $j[i]$ indexes the poststratification cell containing unit $i$ and $f(\hat{\pi}^*_{j[i]})$ is a function of cell-level inclusion probabilities. 
A flexible choice of $f$ guards against misspecification. 
\cite{si_2025} discretizes the estimated inclusion probabilities and includes a varying intercept across the $G$ discrete values along with a continuous term.
The resulting model is
\begin{align*} 
y_{i} \mid \mathbf{x}_{j[i]}, \pi^*_{j[i]}, \Pi_{g[j]}, \eta_{d[i]}  &\sim \mathrm{Bernoulli}(p_{i}),\\  \mathrm{logit}(p_i)  &= \mathbf{x}_{j[i]}^\top \boldsymbol{\beta}   + \beta_\pi \,\mathrm{logit}(\pi^*_{j[i]})\\
&\quad + \Pi_{g[i]} + \eta_{d[i]},
\end{align*}
where $\beta_\pi$ is a fixed-effect coefficient for the continuous inclusion probability, $\Pi_{g}$ is a random intercept for the discretized inclusion probability bins $g = 1,\dots,G$, and $\eta_{d}$ is the domain-level random effect for $d = 1,\dots,D$.

The inclusion probabilities and outcome model can be estimated jointly in a fully Bayesian procedure \citep{si2015bayesian} or point estimates of the inclusion probabilities can be plugged into the outcome model \citep{si_2025}.
The covariates $\mathbf{x}_j$ do not necessarily have to match $\boldsymbol{\gamma}_j$ from (\ref{eq:mrp_inclusion_model}).
Poststratification then aggregates cell-level predictions using known population cell counts yielding population or domain-level estimates.

When the joint, population distribution of poststratifiers is unavailable, these counts can instead be estimated using the PS.
Nonparametric bootstrap approaches have been proposed to generate synthetic populations as a substitute for the full population frame: \cite{dong2014nonparametric} introduced the weighted finite population Bayesian bootstrap for complex sampling designs, \cite{si2015bayesian} extended this to weighted sampling inference, and \cite{li2024embedded} unified these approaches in the embedded MRP framework.
Using estimated counts will generally increase the variability of domain estimates and  requires a large PS where most population cells are represented \citep{si_2025}.

Like mass imputation, all MRP methods reduce DDC through the specification of an outcome model.
The variants that also model inclusion probabilities additionally reduce DDC in the same manner as the DR methods described in the next section. 
However, MRP methods have not been shown to have the doubly robust property.

\subsection{Doubly Robust (DR) Estimators}\label{sec:dr}

DR estimators also directly minimize DDC by specifying both an IPW model for the NPS inclusion weights $\hat{w}_i^*$ and a superpopulation model $\hat{m}(\mathbf{x}_i)$ for the outcome. 
In order for a method to be considered doubly robust, it must satisfy the property that it is consistent so long as either the IPW model or the superpopulation model is correctly specified, but not necessarily both \citep{chen_et_al2019,wu2022}.  
If $\pi_i^* \hat{w}_i^* \approx 1$, then the covariance in (\ref{eq:cov_decomposition}) vanishes regardless of $\hat{m}$.
If $\hat{m}(\mathbf{x}_i) \approx \mathbb{E}(y_i \mid \mathbf{x}_i)$, the residuals $e_i$ carry no selection signal and the covariance vanishes regardless of the IPW model.
Two commonly used DR estimators of the finite-population mean are
\begin{align*}
\hat{\bar y}_{\mathrm{DR1}}
&=
\frac{1}{N}\left(
\sum_{i\in S^*}{\hat{w}_i^*}(y_i-\hat{m}(\mathbf{x}_i)) +  \sum_{i\in S}w_i\hat{m}(\mathbf{x}_i)\right) \nonumber \\
\hat{\bar y}_{\mathrm{DR2}}
&=
\frac{1}{\sum_{i\in S^*}\hat{w}_i^*}
\sum_{i\in S^*}\hat{w}_i^*(y_i-\hat{m}(\mathbf{x}_i))\\
 &\quad +  \frac{1}{\sum_{i\in S}{w}_i}\sum_{i\in S}w_i\hat{m}(\mathbf{x}_i), \nonumber
\end{align*}
The first estimator requires $N$ to be known while the second replaces it with estimated values.

The covariates may differ between the IPW and superpopulation models or can be shared.
Usually these methods assume the response variable is available only in the NPS, not the PS, although \cite{seaman_2025} extend the method of \cite{chen_et_al2019} to the setting where the response is recorded in both samples.
Analytic variance estimation for DR estimators is generally difficult \citep{wu2022}, so bootstrap procedures are common, but a commonly used analytic estimator for the variance is provided by \cite{kim_haziza_2014}.

The application of DR methods to SAE remains relatively underdeveloped. 
\cite{schirripa_spagnolo2024} extend the DR framework to SAE by incorporating area-level random effects in both the IPW model and the superpopulation model with variance estimated via bootstrap.
However, their method applies only to the specific setting where the probability sample and non-probability sample overlap and the analyst knows which units belong to this intersection.

\subsection{Aggregate-level Combination}\label{sec:vsw}

Unlike superpopulation methods which model the outcome variable, aggregate-level combination methods model the bias and form a weighted average of PS and NPS estimates at the domain level.
The estimated mixing weight adapts to the estimated NPS bias and can be viewed as an IPW weighting scheme, where every unit within a domain is assigned the same weight.

\cite{aliste2025combining} (hereafter VSW) propose a method for combining probability and nonprobability samples at an aggregated level when the data are categorical, taking values $c\in\{1,\ldots,C\}$.
Let $y_{dc}$ denote the true population proportion of category $c$ in domain $d$, and let $\hat{\bar{y}}_{dc}$ and $\hat{\bar{y}}^{*}_{dc}$ be the corresponding direct estimators from a PS and NPS, respectively.
The combined estimator is defined as the weighted average
\[
\hat y_{dc}
= W_{dc}\hat{\bar{y}}_{dc} + (1-W_{dc})\hat{\bar{y}}^*_{dc},
\]
where the weight $W_{dc}\in[0,1]$ is chosen to minimize the expected mean squared error (EMSE) of $\hat{\bar {y}}_{dc}$ under a model for the bias $b_{dc} = \mathbb{E}_{\mathrm{des}}(\hat{\bar{y}}^{*}_{dc}) - y_{dc}.$
The authors posit the following random-effects model for the bias,
\[
\mathbb{E}_b(b_{dc})=b_c,\quad
\mathrm{Var}_b(b_{dc})=s^2,
\quad
\sum_{c=1}^C b_c=0,
\]
which assumes category-specific but domain-invariant average bias.
Here $\mathbb{E}_{\mathrm{des}}(\cdot)$ denotes expectation with respect to repeated sampling under both the PS sampling design and the (unknown) NPS inclusion mechanism, while $\mathbb{E}_b(\cdot)$ denotes expectation under the posited random-effects model for the NPS bias.
Under this model, the EMSEs of the two estimators are
\begin{align*}
\mathrm{EMSE}(\hat{\bar{y}}^{*}_{dc})
&= b_c^2 + s^2 + \frac{v_{dc}}{n^{*}_d-1}\\
\mathrm{EMSE}(\hat{\bar{y}}_{dc})
&= \frac{1}{n_d}\Bigl\{ \frac{n_d^{*}}{n_d^{*}-1}v_{dc} + b_c\bigl(2\mathbb{E}(\tilde y_{dc})-1\bigr) - b_c^2 + s^2 \Bigr\},
\end{align*}
where $v_{dc}=\mathbb{E}_b\mathbb{E}_{\mathrm{des}}[\hat{\bar{y}}^{*}_{dc}(1-\hat{\bar{y}}^{*}_{dc})]$ and $\tilde y_{dc}=\mathbb{E}_{\mathrm{des}}(\hat{\bar{y}}^{*}_{dc})$.
The PS EMSE expression involves NPS quantities because the joint expectation $\mathbb{E}_b\mathbb{E}_{\mathrm{des}}$ integrates over both the sampling design and the posited distribution of NPS bias.
The optimal weight is then
\[
W_{dc}
=
\frac{\mathrm{EMSE}(\hat{\bar{y}}^{*}_{dc})}
{\mathrm{EMSE}(\hat{\bar{y}}^{*}_{dc})+\mathrm{EMSE}(\hat{\bar{y}}_{dc})}.
\]

The optimal weight $W_{dc}$ governs the proportion that the PS and NPS contribute to the final estimate, respectively.
As the estimated NPS bias $b_c^2 + s^2$ grows relative to PS sampling variance, $W_{dc}$ increases towards $1$ and the combined estimator shifts toward the PS.
In the limit of severe NPS bias, the estimator converges to the PS direct estimator.
A notable feature of the approach is that it requires no individual-level modeling of $\pi_i^*$ or specification of an outcome regression.
The only inputs are domain-by-category summary statistics from each sample, limiting it to settings where the target quantity is expressible in terms of cell proportions.

A related hybrid approach is proposed by \cite{ciginas_evaluating_2025}.
Unlike VSW, it replaces the raw NPS direct estimator with an IPW-corrected NPS estimator and then combines that estimator with the PS estimator.
Its mixing coefficient depends on the estimated MSE of the IPW component, including a bias term estimated from the discrepancy between the IPW and PS estimators, which makes the method less sensitive to propensity-score misspecification. Additionally, unlike VSW, the method provides a variance estimate for the combined estimator.

\subsection{NPS-informed Priors (NIP)}\label{sec:nps-prior}
Bayesian methods offer a means of integrating information from an NPS through the specification of prior distributions informed by the NPS data.
We refer to this class of methods as NPS-informed prior (NIP) methods.
Like superpopulation methods (Section~\ref{sec:sp}), this class of methods fits a regression model to NPS units and uses the resulting coefficient estimates to obtain estimates of finite population quantities.

\cite{sakshaug_2019} propose a linear regression model with conditionally conjugate priors on the regression coefficients and error variance.
They note a reduction in variance and therefore MSE compared to estimates obtained using only the probability sample.
\cite{wisniowski_2020} extend this approach by developing a fully conjugate regression model.
\cite{nandram2021bayesian, nandram_2024} extend these methods to small-area estimation by incorporating the modeling framework of \cite{battese1988}.
Additionally, they analyze the bias introduced by the nonprobability sample using the framework of \cite{meng2018} and provide numerical evidence that use of the nonprobability sample in the prior introduces less bias than using it in the likelihood.
\cite{you2021approximate} adapt these methods to estimate a finite population mean and show that the proposed estimator achieves lower bias and MSE than the Horvitz--Thompson estimator.
They also discuss properties of the proposed estimator in the contexts of both ignorable and non-ignorable designs.
\cite{salvatore_2023} adapt these methods to binary data by developing a logistic regression model.

We focus on the method of \cite{salvatore_2023}, with a proposed extension that better fits the SAE setting under consideration in the present work. 
The authors incorporate the survey weights as a covariate and estimate the NPS weights via raking.
They only consider inference, not SAE.
To extend to SAE, we instead perform post-stratification to produce population estimates.
We follow \cite{you2021approximate} and use the PS weights as exponents in a pseudolikelihood.

\cite{salvatore_2023} consider several families of prior distributions on the regression coefficients $\boldsymbol{\beta}$. 
We focus on the \textit{power prior} \citep{chen_1999, chen_2000, ibrahim_2000}.
The power prior based on the NPS takes the form
\begin{equation}
\label{eq:powerprior}
    p(\boldsymbol{\beta} \mid \mathbf{y}^*, \mathbf{X}^*) \propto \mathcal{L}(\boldsymbol{\beta} \mid \mathbf{y}^*, \mathbf{X}^*)^a p(\boldsymbol{\beta}).
\end{equation}
To set $a$, the authors fit a logistic regression to the PS and obtain the MLE $\hat{\boldsymbol{\beta}}$ of $\boldsymbol{\beta}$. 
Analogously, they obtain the MLE $\hat{\boldsymbol{\beta}}^\star$ from the NPS.
They then set $a$ to the $p$-value from a Hotelling $T^2$ test of equality between the two estimates.
This test statistic is
\begin{equation*}
    T^2 = \left(\hat{\boldsymbol{\beta}} - \hat{\boldsymbol{\beta}}^*\right)^\top \left[\hat{\mathrm{Cov}}\left(\hat{\boldsymbol{\beta}}\right) + \hat{\mathrm{Cov}}\left(\hat{\boldsymbol{\beta}}^*\right)\right]^{-1} \left(\hat{\boldsymbol{\beta}} - \hat{\boldsymbol{\beta}}^*\right).
\end{equation*}
Values of $a$ near zero indicate disagreement and downweight the NPS contribution to the power prior while values near one allow the NPS to substantially weight the power prior. 
We can view this as the power prior incorporating the NPS data only to the degree that PS--NPS agreement suggests transportability is plausible. 

The full power prior-based model is
\begin{equation*}
    \begin{aligned}
        \mathcal{L}(\boldsymbol{\beta} \mid \mathbf{y}, \mathbf{X}, \mathbf{y}^*, \mathbf{X}^*) & \propto \left[\prod_{i \in S} \frac{\exp(\mathbf{x}_i^\top \boldsymbol{\beta} + \eta_{d[i]})^{w_i y_i}}{[1 + \exp(\mathbf{x}_i^\top \boldsymbol{\beta} + \eta_{d[i]})]^{w_i}} \right]\\
        &\quad\times\left[\prod_{i \in S^*} \frac{\exp(\mathbf{x}_i^\top \boldsymbol{\beta} + \eta_{d[i]})^{a y_i}}{[1 + \exp(\mathbf{x}_i^\top \boldsymbol{\beta}+ \eta_{d[i]})]^{a}} \right] \\
        \boldsymbol{\beta}  & \sim \mathcal{N}(0, \sigma_\beta^2\boldsymbol{I}) \\
        \eta_1, \ldots, \eta_D \mid \sigma_\eta^2 & \overset{\text{iid}}{\sim} \mathcal{N}(0, \sigma_\eta^2).
    \end{aligned}
\end{equation*}
The values of $\sigma_\beta$, the prior on $\sigma_\eta$, and the computational implementation are described in Section~\ref{sec:sim-methods}.

In their empirical study, \cite{salvatore_2023} observe that estimates become prior-invariant at PS sizes greater than 500. 
This occurs because the $p$-value approaches 0 as the PS size increases.
Thus, given a large PS, the NPS will never contribute to the power prior.
Another drawback of using a $p$-value in the power prior is that, under the null hypothesis of the Hotelling $T^2$ test (i.e., no difference between the PS and NPS coefficients), the $p$-value follows a $\mathrm{Unif}(0, 1)$ distribution.
Therefore, the expected value of $a$ is 0.5.
But, when the PS and NPS are perfectly compatible, we expect $a$ to concentrate around a value much closer to 1.

SAE settings typically have PS sample sizes much greater than the cut-off noted by \cite{salvatore_2023}. 
To better study the effect of the NIP methods in an SAE context, we instead propose replacing the $p$-value in (\ref{eq:powerprior}) with a smooth transformation of the $T^2$ test statistic
\begin{equation*}
    a = \exp\left(\frac{\log(0.9)}{E_{H_0}[T^2]} T^2\right).
\end{equation*}
This quantity is calibrated so that $a = 0.9$ when $T^2$ equals its expectation under $H_0$. 
In Appendix~\ref{app:exp-link} we show that, under $H_0$, $T^2$ is asymptotically $\chi^2_q$, where $q$ is the dimension of $\boldsymbol{\beta}$. 
Thus, $E_{H_0}[T^2] = q$ with a null distribution that is invariant to the sample size.
The resulting null mean of $a$ is approximately $0.9$, in contrast to the $p$-value, which is uniform. 
When the samples are incompatible, $a$ still tends to zero as evidence of the discrepancy accumulates, but much more slowly than the $p$-value does.

\section{Simulation Study}\label{sec:sim}

We conduct a design-based simulation study using ACS PUMS data to evaluate the methods described in Section~\ref{sec:overview}.
The estimators draw on different combinations of the probability sample, the nonprobability sample, and population-level covariate information. 
Two PS-only estimators are included as benchmarks.
All estimates are compared against a known finite-population truth constructed from ACS PUMS.
Code and computational details are provided in the Reproducibility section.

\subsection{Population and Response Variable}

We use the 2023 ACS PUMS for California as our study population, from which we draw probability and non-probability samples.
We treat the PUMA (Public Use Microdata Area) level means computed from these data as ground-truth values that allow us to assess how well each method estimates a population quantity from sample data.
The full microdata comprise $N = 1{,}475{,}631$ individual records, partitioned across $D = 281$ Public Use Microdata Areas (PUMAs).
We refer to the microdata values as our ``oracle finite population" to distinguish them from the full U.S. population, of which the ACS is a representative sample.

Our response variable is means-tested public health coverage (Medicaid and related programs; \textsc{pubcov} in the ACS).
This does not include Medicare, which is available to nearly all adults aged 65 and older regardless of income.
The population prevalence is 60.3\%.
Each individual is also associated with three covariates used in all downstream models: age group (\textsc{age}, binned into four categories), race/ethnicity (\textsc{race}, binned into five categories), and sex (\textsc{sex}, two categories).
We assume the joint population counts for all of these categories are available as this is often the case in practice. 

\subsection{Sampling Design}

For $B = 100$ simulation replicates we independently draw both a probability sample and a nonprobability sample from the ACS ``population."
Both samples are drawn with unequal-probability designs in which inclusion probabilities are proportional to a size variable, but with a different size variable for the PS and NPS, respectively.
This sampling step is carried out with the R \textit{sampling} library \citep{sampling_library}.

The NPS design is intended to mimic a biased opt-in survey, in which units self-select with unknown inclusion probabilities.
Accordingly, the NPS inclusion probabilities are treated as unknown and the corresponding survey weights are never used in the analysis.
As such, when we draw the NPS, each individual $i$ is assigned a size variable
\[
  q_i \propto \exp\left(
    t_{\mathrm{pwgt}}\,z_{\mathrm{pwgt},i}
    + t_{\mathrm{pov}}\,z_{\mathrm{pov},i}
  \right),
\]
where $z_{\mathrm{pwgt},i}$ and $z_{\mathrm{pov},i}$ are standardized \textsc{pwgtp} (ACS person-level design weight) and \textsc{povpip} (income-to-poverty ratio), and $(t_{\mathrm{pwgt}}, t_{\mathrm{pov}})$ are tuning parameters.
Individual inclusion probabilities are set proportional to $q_i$ and scaled to achieve a target sampling fraction of $f^* = 0.05$ yielding an expected NPS sample size of $74,000$.
Crucially, neither \textsc{pwgtp} nor \textsc{povpip} is used as a model covariate in the simulation study estimators, so the selection mechanism operates through factors that will not be directly conditioned on.

By varying the tuning parameters $(t_{\mathrm{pwgt}}, t_{\mathrm{pov}})$ we obtain the three settings shown in Table~\ref{tab:ddc_settings}.
These settings are calibrated to reflect the range of biases in non-probability samples documented by \cite{pew_2023}.
The ``favorable" setting approximates a low-bias opt-in panel, the ``typical" setting matches the average absolute error across their benchmarks, and the ``extreme" setting reflects the worst-case errors observed.

\begin{table}[ht]
\centering
\caption{NPS bias configurations, averaged over 20 calibration runs.
     Bias is the difference between the unweighted NPS sample mean and the population mean of \textsc{pubcov} (60.3\%), in percentage points.
     $n_{\mathrm{eff}}$ is the bias-aware effective sample size (Section~\ref{sec:ddc}).
\label{tab:ddc_settings}
    }
\begin{tabular}{lccccr}
  \toprule
  Setting & $t_{\mathrm{pwgt}}$ & $t_{\mathrm{pov}}$
          & DDC ($\rho_{r^*,y}$) & $n_{\mathrm{eff}}$ & Bias (pp) \\
  \midrule
  Extreme   & $0.10$ & $-1.52$ & $-0.089$ & $\approx 7$     & $-19.0$ \\
  Typical   & $0.10$ & $-0.41$ & $-0.029$ & $\approx 64$    & $-6.1$ \\
  Favorable & $0.10$ & $-0.22$ & $-0.014$ & $\approx 255$   & $-3.0$ \\
  \bottomrule
\end{tabular}
\end{table}

\begin{figure*}[h]
\centering
\includegraphics[width=.9\textwidth]{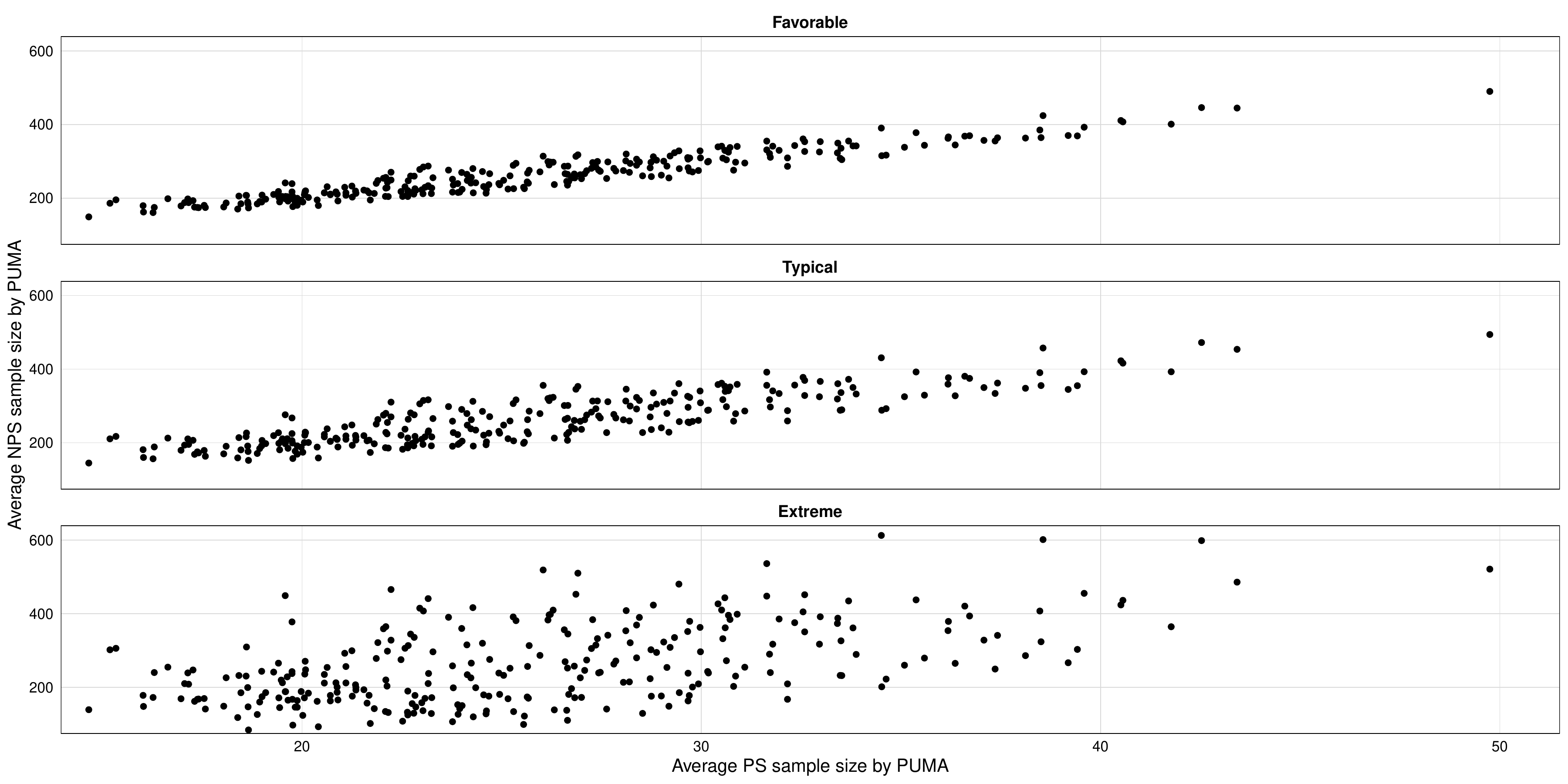}
\caption{Scatterplot of average PS sample size against average NPS sample size by PUMA for all three DDC settings. 
The probability sample is the same across all three DDC settings. 
Only the NPS varies.}
\label{fig:sample_size_comparison}
\end{figure*}

The scatterplot in Figure~\ref{fig:sample_size_comparison} plots the average PS sample size across all simulations for each PUMA against the average NPS sample size.
We see that the NPS samples are much larger, typically by a factor of ten.
The NPS sample size also becomes more variable as the DDC increase.

\begin{figure*}[!h]
\centering
\includegraphics[width=.7\textwidth]{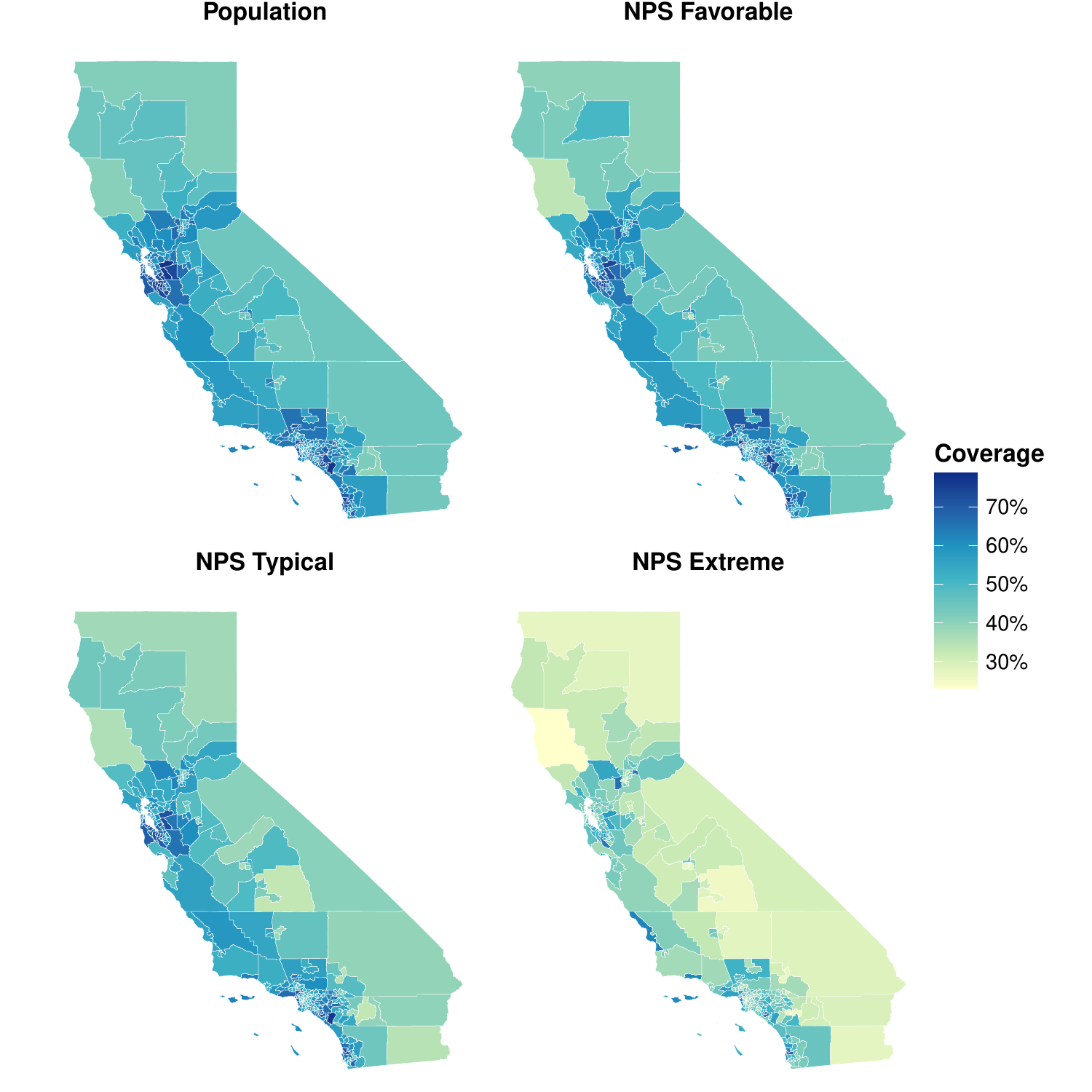}
\caption{Mean \texttt{PUBCOV} in California by PUMA. The facets compare the true population proportion (top left) against sample means from three different NPS samples, one from each DDC configuration (favorable, typical, and extreme).
}
\label{fig:nps_maps}
\end{figure*}

Figure~\ref{fig:nps_maps} displays the mean response by PUMA for the ACS population and three NPS samples, one for each DDC setting. 
The degree of agreement between the population and the NPS samples diverges as the magnitude of the DDC increases.

By contrast, PS units are sampled proportional to the size variable 
\[q_i \propto \exp(0.05\cdot z_\text{wagp} - 0.2\cdot z_\text{pwgt}),\]
where $z_{wagp}$ is standardized person-level yearly income.
The PS sample is also stratified, since some methods require all domains to have at least $n_d=1.$
Strata are defined by PUMAs and a within-stratum sampling fraction of $f = 0.005$ is targeted, yielding an expected PS total sample size of $7{,}264$.
Design weights are set to inverse inclusion probabilities.
We use Poisson sampling because it provides a realistic, unequal probability sampling design which necessitates the use of survey weights while still being computationally efficient.
Under this design, the realized DDC across replicates is approximately $\mathbb{E}[\rho_{r,y}] = 0.0007$, which indicates $\mathbb{E}[\rho_{r,y}] \lesssim N^{-1/2} \approx 0.0008$.
Thus, the condition  of \cite{meng2018}  that the simulated PS behaves like a true, quality PS is met.
Samples are drawn separately within each PUMA with a minimum target of one draw per PUMA.
Realized sample sizes remain random under Poisson sampling.

\subsection{Methods}\label{sec:sim-methods}

We evaluate estimators drawn from the methodological families reviewed in Section~\ref{sec:overview}.
Table~\ref{tab:methods} summarizes all estimators.
All estimators except HT-PS and VSW use unit-level covariates $\mathbf{x}_i = (\textsc{age}, \textsc{race}, \textsc{sex})^\top$.
We also adapt many estimators to include PUMA random effects to make them suitable for SAE. 
Two estimators use only the PS and serve as benchmarks: the Horvitz--Thompson direct estimator (HT-PS) and a Bayesian unit-level model fit to the PS (BULM-PS) using the pseudolikelihood of Section~\ref{sec:ipw}.
All Bayesian models are fit using \texttt{cmdstanr} with 2 chains, 1{,}000 warmup iterations, and 1{,}000 sampling iterations per chain.

\begin{table*}
\centering
\caption{Summary of estimators grouped by methodological family.
PS, NPS, and Population indicate whether the method uses the probability sample, nonprobability sample, and ACS population frame, respectively.
+Y variants additionally include the response variable in the selection model.
$^\ddag$
Coverage and interval score are not reported for VSW, which does not produce interval estimates. 
}
\label{tab:methods}
\begin{tabular}{lccc} 
\toprule Method & PS & NPS & Population \\ 
\midrule \multicolumn{4}{l}{\textit{Benchmarks}} \\
Direct estimate (Direct) & $\checkmark$ & & \\
Bayesian unit-level model (BULM-PS) & $\checkmark$ & & $\checkmark$ \\
\midrule \multicolumn{4}{l}{\textit{IPW}} \\ 
DE with IPW weights (IPW-DE) & $\checkmark$ & $\checkmark$ & \\
DE with IPW weights + response (IPW-DE+Y) & $\checkmark$ & $\checkmark$ & \\
BULM with IPW weights (IPW-BULM) & $\checkmark$ & $\checkmark$ & $\checkmark$ \\
BULM with IPW weights + response (IPW-BULM+Y) & $\checkmark$ & $\checkmark$ & $\checkmark$ \\
DE with CLIP weights (IPW-DE-CLIP) & & $\checkmark$ & $\checkmark$ \\ 
\midrule \multicolumn{4}{l}{\textit{MRP}} \\
MRP with CLIP and pop. poststrat. (MRP-INT-P) & & $\checkmark$ & $\checkmark$ \\
MRP with IPW weights and pop. poststrat. (MRP-INT-P-IPW) & $\checkmark$ & $\checkmark$ & $\checkmark$ \\ 
\midrule \multicolumn{4}{l}{\textit{Aggregate combination}} \\
VSW$^\ddag$ & $\checkmark$ & $\checkmark$ & \\ 
\midrule 
\multicolumn{4}{l}{\textit{NIP}} \\
NIP (p-value) & $\checkmark$ & $\checkmark$ & $\checkmark$ \\
NIP (exp link) & $\checkmark$ & $\checkmark$ & $\checkmark$ \\
\bottomrule 
\end{tabular}
\end{table*}

\subsubsection{Default priors}
Unless otherwise noted, fixed-effect coefficients priors are taken to be $\boldsymbol{\beta} \sim \mathcal{N}(\mathbf{0}, 9\boldsymbol{I})$ and domain random-effect prior standard deviations are $\sigma_\eta \sim \mathrm{Cauchy}^+(0, 5)$.

\subsubsection{IPW methods}
We implement the IPW variants in Table~\ref{tab:methods} via the weighted logistic regression method presented in \cite{elliott_2017} and described in Section~\ref{sec:ipw}, with the weights used in the logistic regression defined according to (\ref{eq:ipw_weights}) and with race, sex, and age as covariates.
The $+y$ variants additionally include the response variable \textsc{pubcov} in the propensity model, so that the covariates entering the logistic regression become $(\mathbf{x}_i, y_i)$ rather than $\mathbf{x}_i$ alone.

We calculate a direct estimate and BULM estimate using the combined PS and NPS with the estimated weights.  
This requires rescaling the PS design weights $w_i$ and estimated NPS weights $\hat{w}_i^* = 1/\hat{\pi}_i^*$ by
\[
  C_S = \frac{n}{n + n^*} \qquad \text{ and }
  C_{S^*} = \frac{n^*}{n + n^*} \cdot \frac{\sum_{i \in S} w_i}{\sum_{i \in S^*} \hat{w}_i^*}
\]
respectively so that $C_S w_i$ is the final weight for PS units and $C_{S^*}\hat{w}_i^*$ for NPS units.
This normalization ensures that the weighted contribution of each sample reflects its unweighted share in the combined dataset.

We also consider an alternative set of weights derived from the cell-level inclusion probabilities (CLIP) estimated via (\ref{eq:mrp_inclusion_model}).
Unlike the combined-sample IPW weights, these weights require only population cell counts and the NPS. 
We include a random effect for PUMA in the inclusion model.
We use the resulting inclusion probabilities in a MRP-INT-P model (see \ref{sec:mrp_implementation}) and calculate a direct estimator with the inverse of these probabilities used as weights,

Confidence intervals for the IPW-HT estimators are calculated using the R \texttt{survey} library \citep{lumley_2010}.
This does not account for uncertainty from estimating the weights.

We show in Appendix~\ref{app:ipw} that the choice of IPW estimator has little effect on the final estimate.
As such, we consider only IPW weights estimated with weighted logistic regression (\ref{eq:ipw_weights}) and the CLIP weights.

\subsubsection{MRP methods}
\label{sec:mrp_implementation}
We implement MRP-INT-P  with two different inclusion models.
The first uses CLIPs estimated via (\ref{eq:mrp_inclusion_model}).
For MRP-INT-P (IPW), we pool the PS and NPS, define a source indicator $Z_i$ equal to $1$ for NPS records and $0$ for PS records, and fit a weighted logistic regression
\[
  \Pr(Z_i = 1 \mid \mathbf{x}_i)
  = \operatorname{logit}^{-1}\!\bigl(\alpha + \mathbf{x}_i^\top \boldsymbol{\beta} \bigr),
\]
with unit weight $1$ for NPS records and PS design weights for PS records.
Predicted inclusion probabilities are truncated away from $0$ and $1$, then attached to the corresponding \textsc{age}$\times$\textsc{sex}$\times$\textsc{race}$\times$\textsc{puma} cells.
The CLIP logits enter the outcome model through a linear term $\beta_\pi\,\mathrm{logit}(\hat\pi^*_j)$, and a second term is formed by rounding $\hat\pi^*_j$ to two decimal places and assigning a random intercept to each resulting bin.

Following \citep{si_2025} our MRP-INT-P implementation utilizes a two-stage procedure, where the posterior median estimates of the CLIPs are plugged into the MRP outcome model.
Poststratification cells are defined by the full cross-classification of \textsc{age}, \textsc{sex}, \textsc{race}, and PUMA and final estimates are aggregated to the PUMA level.

\subsubsection{Aggregate-level combination}
We implement the VSW estimator of \cite{aliste2025combining} (Section~\ref{sec:vsw}), which does not produce interval estimates.
Coverage and interval score are not reported for this method.

\subsubsection{NPS-Prior methods}

We implement the NIP p-value and exp-link variants in Table~\ref{tab:methods} as described in Section~\ref{sec:nps-prior}.

\subsection{Evaluation Criteria}\label{sec:eval}

We evaluate each estimator across $B = 100$ independent
simulation replications.
In each replication, all methods are applied to the same pair of PS and NPS draws, and PUMA-level estimates are compared against the known finite-population truth $\theta_d = N_d^{-1}\sum_{i \in U_d} y_i$ for domain $d \in \{1, \ldots, D\}$.

Let $\hat{\theta}_d^{(b)}$ denote the point estimate for domain $d$ in replication $b$, and let $[L_d^{(b)},\, U_d^{(b)}]$ denote the corresponding $95\%$ interval estimate.
We report three metrics, each averaged over domains and replications.
\begin{enumerate}
\item Average MSE
\begin{equation*}
  \mathrm{MSE}
  = \frac{1}{B} \frac{1}{D}
    \sum_{b=1}^{B} \sum_{d=1}^{D}
    \bigl(\hat\theta_d^{(b)} - \theta_d\bigr)^{2}.
  \label{eq:mse}
\end{equation*}
\item Mean coverage
\begin{equation*}
  \mathrm{Coverage}
  = \frac{1}{B} \frac{1}{D}
    \sum_{b=1}^{B} \sum_{d=1}^{D}
    \mathbf{1}\!\bigl(L_d^{(b)} \le \theta_d \le U_d^{(b)}\bigr).
  \label{eq:coverage}
\end{equation*}
\item Mean interval score (IS)
\begin{equation*}
  \overline{\mathrm{IS}}_{0.05}
  = \frac{1}{B}\frac{1}{D}\sum_{b=1}^B\sum_{d=1}^D \mathrm{IS}_{0.05}(d,b)
  \label{eq:is}
\end{equation*}
where 
\begin{equation*} 
\begin{aligned} 
\mathrm{IS}_\alpha(d,b) = {}& \bigl(U_d^{(b)} - L_d^{(b)}\bigr)\\ 
&+ \frac{2}{\alpha}\bigl(L_d^{(b)}-\theta_d\bigr)\mathbf{1}\!\bigl(\theta_d < L_d^{(b)}\bigr)\\ 
&+ \frac{2}{\alpha}\bigl(\theta_d - U_d^{(b)}\bigr) \mathbf{1}\!\bigl(\theta_d > U_d^{(b)}\bigr). 
\end{aligned}
\end{equation*}
is the interval score of \cite{gneiting_strictly_2007}, a proper scoring rule that rewards narrow intervals and penalizes those failing to cover the truth; smaller values indicate better performance.
\end{enumerate}

\subsection{Results}\label{sec:results}
\begin{figure*}[th]
\centering
\includegraphics[width=\textwidth]{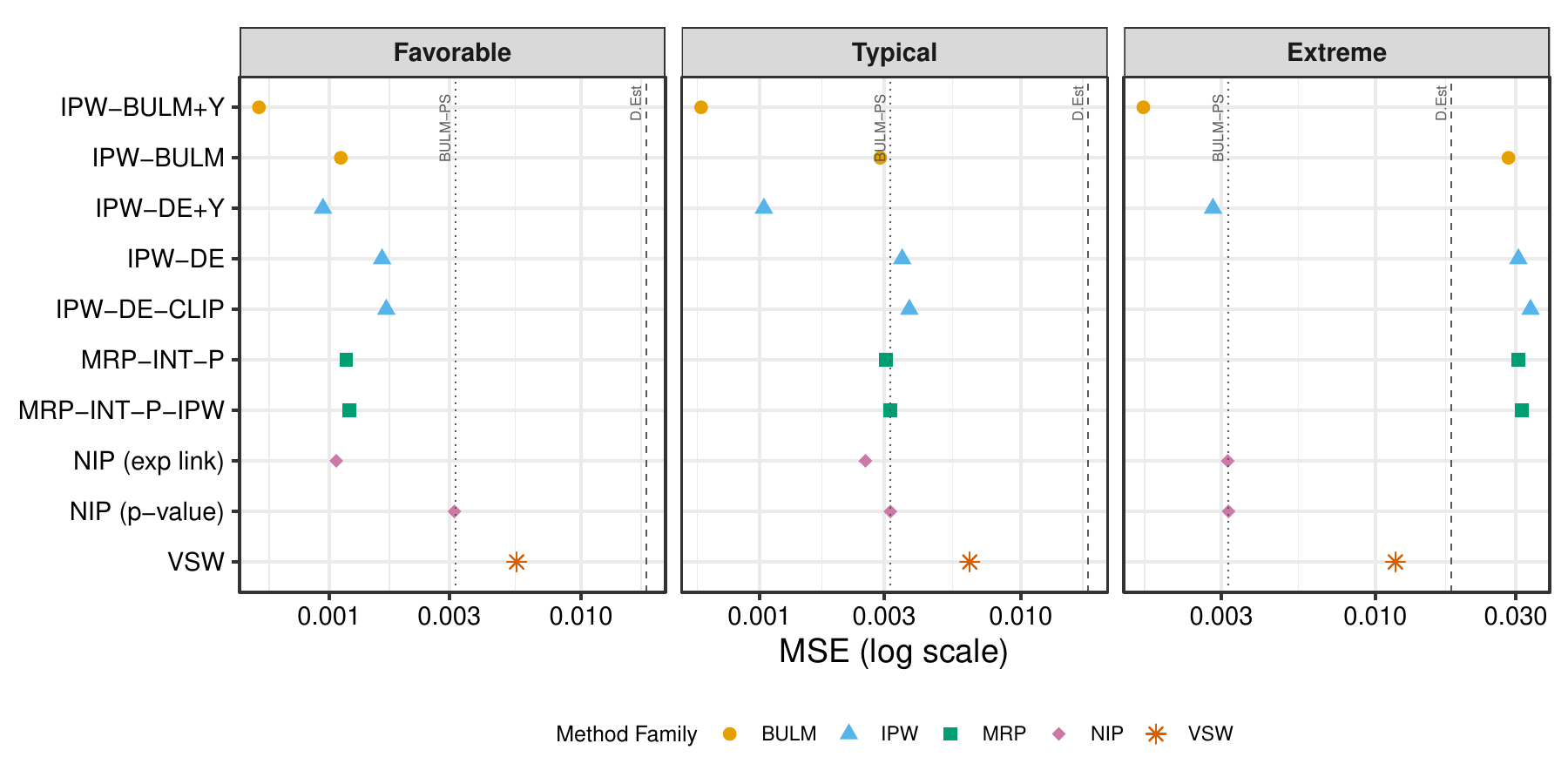}
\caption{MSE on log scale for each method faceted by DDC setting.
Dashed vertical lines indicate PS-only benchmarks.}
\label{fig:mse}
\end{figure*}
\begin{figure*}[th]
\centering
\includegraphics[width=\textwidth]{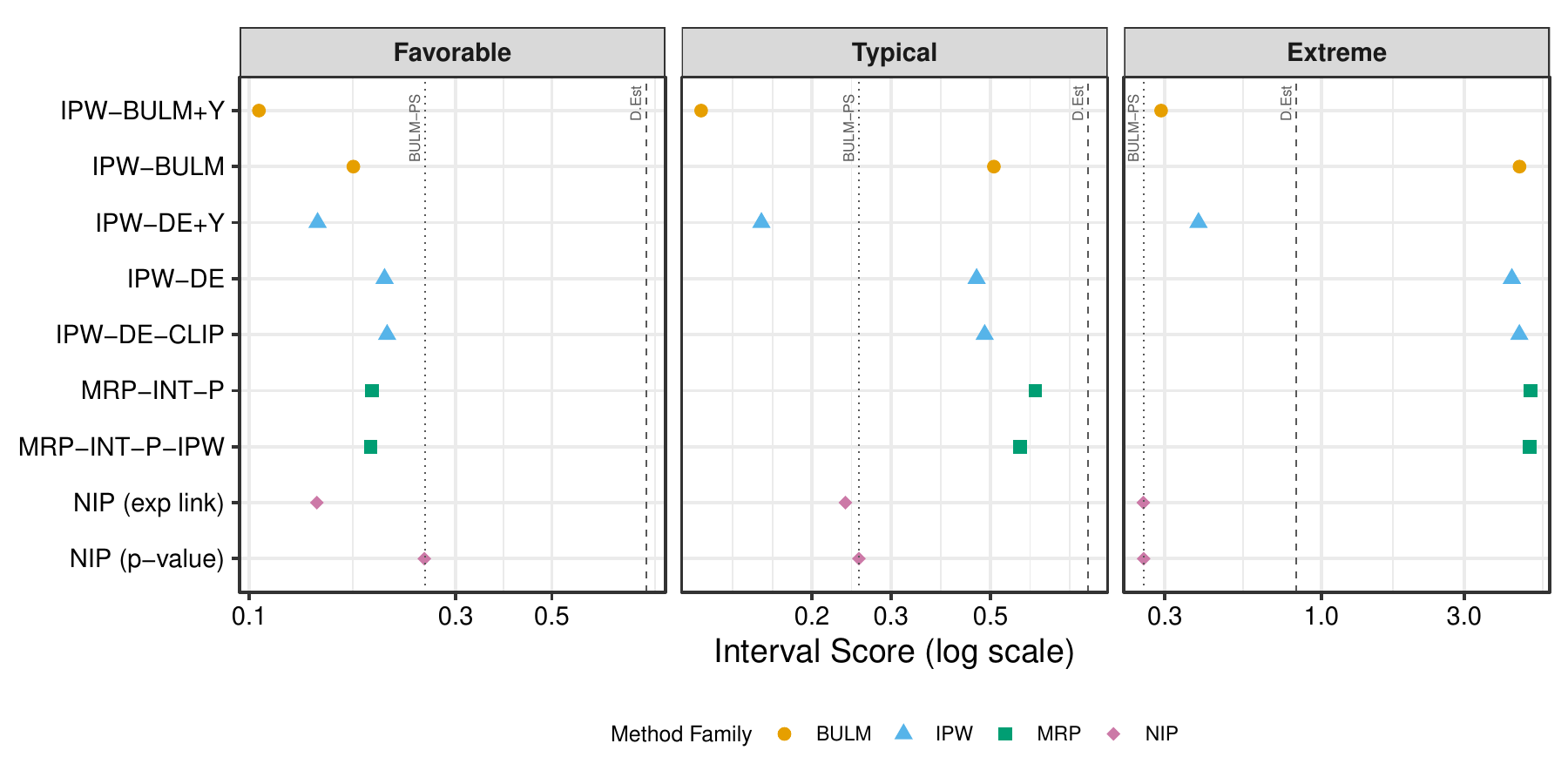}
\caption{
Interval score on log scale for each method, faceted by DDC setting.
Lower values indicate better performance.
Dashed vertical lines indicate PS-only benchmarks.
}
\label{fig:is}
\end{figure*}
\begin{figure*}[th]
\centering
\includegraphics[width=\textwidth]{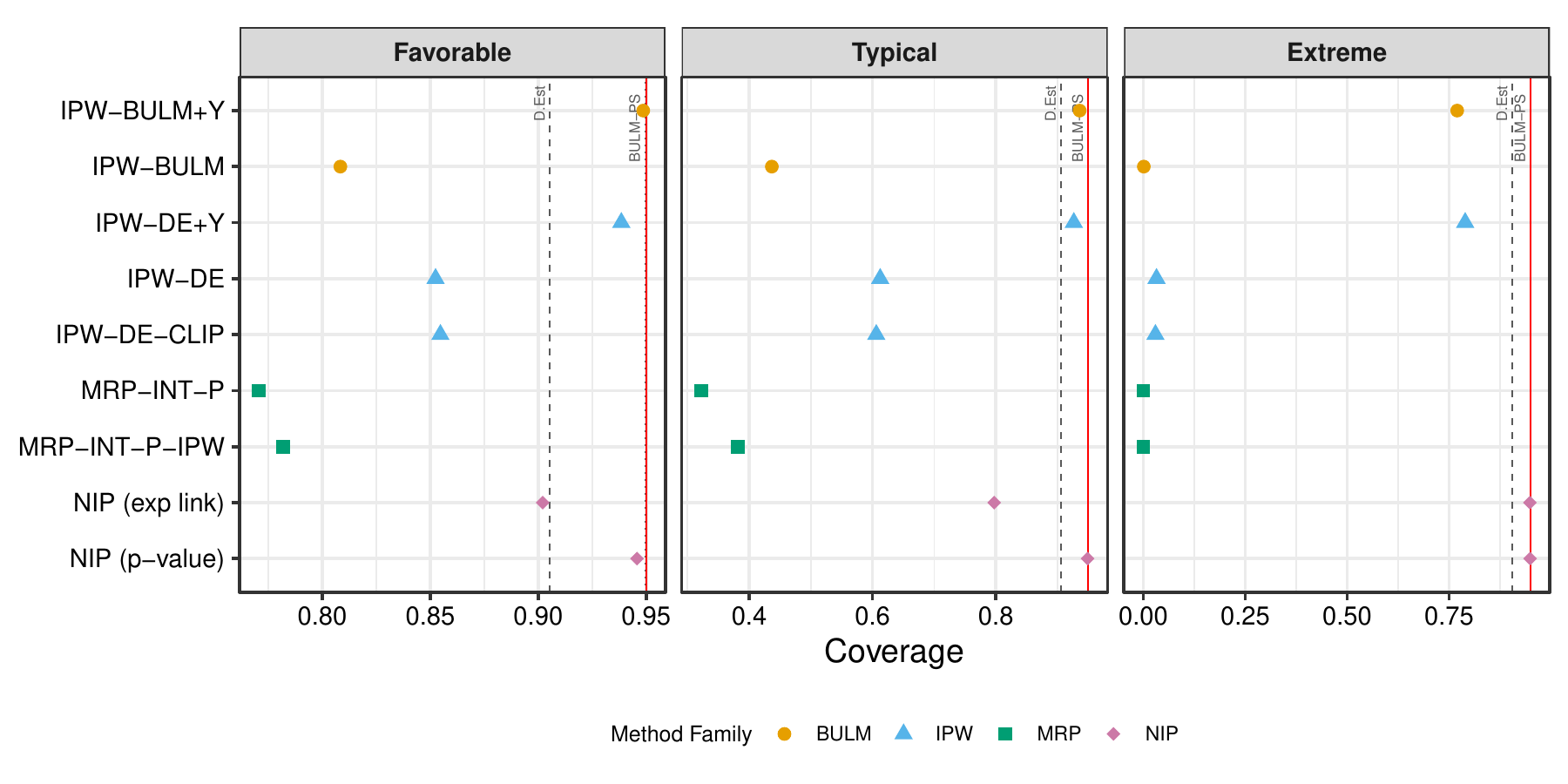}
\caption{Mean interval coverage for each method faceted by DDC setting.
Nominal coverage is 95\%, marked by a solid line.
Dashed vertical lines indicate PS-only benchmarks.}
\label{fig:coverage}
\end{figure*}

Figures~\ref{fig:mse}--\ref{fig:coverage} summarize the results of averaging MSE, IS, and interval coverage across all simulation runs and PUMAs for each method considered.
MSE and IS are plotted on log scales. 
The full numerical results are included as tables in Appendix~\ref{app:results}.

To gauge whether NPS data is helpful in each setting, we include two baseline methods that use only PS data.
The Horvitz--Thompson direct estimator baseline uses only the survey weights and observed responses from the PS without any additional population information.
The BULM-PS baseline is the basic unit-level model (\ref{eq:basic_unit_level}) fit to PS data only, with estimates generated through poststratification.
We expect the latter baseline to perform better, since it relies on auxiliary information and borrows strength through covariates and a PUMA random effect.
In all figures, the two benchmark method values are marked by dashed reference lines. 

We first observe that the methods that include the response variable in a propensity score model (IPW-DE+Y and IPW-BULM+Y) perform the best in terms of MSE.
They attain lower MSE than the benchmarks and all other methods across all DDC settings.
This is most evident for IPW-BULM+Y, which consistently improves on IPW-BULM in all DDC settings, with the gap widening as the DDC setting becomes more difficult.
The same pattern holds for IPW-DE versus IPW-DE+Y.
We emphasize that the availability of the response in the PS is a special case, since NPS analyses are often motivated by the fact that the response is not available in the PS.
Nonetheless, our results indicate that when the response is available in the PS it should be used in conjunction with the NPS data.
Although the IPW-DE methods perform less well than the IPW-BULM methods, the former do not rely on population information and are therefore more generally applicable.

Another important special case to highlight is the setting in which no PS is available but population auxiliary information is. 
We note that IPW-DE-CLIP and MRP-INT-P use only NPS data and population cell counts.
IPW-DE-CLIP performs similarly on all metrics to IPW-DE, which uses weights estimated from combined PS and NPS data. 
Likewise, MRP-INT-P using the CLIPs performs similarly to MRP-INT-P-IPW. 
In fact, both methods based on CLIPs perform nearly as well as the BULM-PS baseline in the favorable and typical settings despite not using PS design weights.
These results suggest that, outside of the extreme DDC setting, lack of a PS is not necessarily prohibitive for an analysis when population cell counts are available.

Several other general patterns are visible in the results. 
In the favorable setting, most methods outperform both the direct estimate and BULM-PS baselines on MSE and IS. 
This is expected, since in the favorable setting the NPS data quality is close to that of the PS, and the effective sample size is large.
This advantage diminishes in the typical setting, where only the +Y variants and NIP exp-link model maintain a clear edge over BULM-PS.
In the extreme setting, only the +Y methods improve on the BULM-PS baseline in terms of MSE, and only the NIP methods match the baseline on IS.
Many of the methods in the extreme case even underperform the direct estimate on all metrics, and interval coverage collapses nearly to zero for all but the +Y and NIP methods.

The NIP prior methods are remarkably robust to differences in DDC setting.
The exp-link variant in particular performs well.
The $p$-value variant behaves nearly identically to the BULM-PS model on all metrics and across DDC settings.
This occurs because the PS sample size eliminates the contribution of the NPS to the power prior as discussed in Section~\ref{sec:nps-prior}.
By contrast, the exp-link variant improves on both the $p$-value variant and the BULM-PS model in terms of MSE and IS under the favorable and typical DDC settings.
The improvement is greatest in the favorable setting, where the NPS contributes more to the power prior. 
In the extreme setting, where the NPS is of lower quality, the performance matches the BULM-PS.
In this sense, the NIP exp-link approach adaptively interpolates between borrowing more information from the NPS when it is valuable and reverting to behavior similar to BULM-PS when the additional information is not useful.
This comes at the cost of worse CI coverage in the non-extreme DDC settings, whereas the $p$-value variant and BULM-PS generally attain nominal coverage rates. 

The VSW estimator does not produce interval estimates and its MSE exceeds that of the BULM-PS in  all three DDC settings. 
However, it stands out for outperforming the direct estimator while requiring no population-level information.
Therefore, the method is quite general as it applies in cases where population information is unavailable.
This type of method may prove to be a fruitful avenue of future research. 

In practice, data analysts are unlikely to know the DDC setting in which they are operating.
Hence, the results of the simulation study imply that caution must be taken before deciding to use NPS data in an analysis. 
When the response variable is available in the PS, then methods that leverage this property should be used.
If population information is available this can take the form of a unit-level model, otherwise a direct estimator.
When the response is not available in the PS and there is doubt about the quality of the data, NIP methods are a robust choice.
If point estimation is the goal, IPW-BULM+Y displays the best MSE at the cost of undercoverage, whereas the NIP methods attain closer to nominal coverage.

\section{Discussion}
\label{sec:discussion}

This paper reviews recent developments in NPS methodology and details how they can be extended and applied in the context of SAE.
The overview and simulation study are framed through the DDC, an important conceptual tool provided by \cite{meng2018}.
We conduct an empirical simulation study that compares the major types of NPS methods along with some newer approaches across varying DDC settings.
The DDC framework provides a novel basis for comparing the methods, assessing their trade-offs, and assessing when NPS data are likely to improve small area estimates.

NPS data are attractive because they increase the apparent sample size and may appear to provide information where PS data is limited. 
However, our simulation study reaffirms the cautionary Big Data Paradox of \cite{meng2018}: more data does not necessarily yield better estimates. 
If NPS data are used carelessly, they produce misleading estimates.

Our results suggest several important directions for future research.
First, practical diagnostics for NPS data quality are needed.
The relative performance of the methods depends greatly on the NPS data quality as measured by DDC, which may not be identifiable from a single sample.

Second, we observe that some methods perform well when NPS data quality is high, but break down tremendously when DDC increases, while other methods sacrifice some performance in favorable settings but remain more stable under increasing DDC.
The performance of the NIP methods suggest that robustness to high DDC is a desirable property for future methodological development. 
In the absence of data quality diagnostics, it will be desirable to search for additional methods that are robust to DDC.

\section*{Reproducibility}

The R code for reproducing all results in this paper is available in the supplementary materials and at \url{https://github.com/dvedensk/NPS_SAE/}.

\begin{acks}[Acknowledgments]
This article is released to inform interested parties of ongoing research and to encourage discussion. The views expressed on statistical issues are those of the authors and not those of the U.S. National Science Foundation or U.S. Census Bureau. 
This research was supported in part by the Natural Sciences and Engineering Research Council of Canada and partially supported by the NSF under NSF grants NCSE-2215168 and NCSE-2215169.
Daniel Vedensky is the corresponding author.
\end{acks}

\appendix
\setcounter{figure}{0}
\setcounter{table}{0}
\renewcommand{\thefigure}{A\arabic{figure}}
\renewcommand{\thetable}{A\arabic{table}}

\section{Details on the NIP exponential link method}
\label{app:exp-link}

Here we provide the details on our extension to the method proposed by \cite{salvatore_2023} that we have modified for the large-sample SAE setting. 
We focus on the version of their method using a power prior which takes the form 
$$\pi(\boldsymbol{\beta} \mid D_{\mathrm{NPS}}, a) \propto L(\boldsymbol{\beta}; D_{\mathrm{NPS}})^a \, \pi_0(\boldsymbol{\beta}),$$ where $a \in [0, 1]$ controls how much the NPS likelihood informs the prior.
\cite{salvatore_2023} set $a$ to the $p$-value of a Hotelling $T^2$ test comparing the PS and NPS regression coefficients. 
As noted in Section~\ref{sec:nps-prior}, this choice has two drawbacks. 
First, under $H_0$ the $p$-value is $\mathrm{Unif}(0,1)$, so even a perfectly compatible NPS receives a weight averaging $0.5$. 
Second, under any fixed discrepancy the $p$-value vanishes as the sample sizes grow, so at larger sample sizes the NPS is discarded regardless of bias.

Our modification replaces the $p$-value transform with
\begin{equation*}
    a = \exp(-c\, T^2), \qquad c = -\log(a_0)/q,
\end{equation*}
where $q = \dim(\boldsymbol{\beta})$ and $a_0 \in (0,1)$ is a calibration target. 
We use $a_0 = 0.9$ throughout.

Under $H_0$ (no difference between the PS and NPS coefficients), both $\hat{\boldsymbol{\beta}}$ and $\hat{\boldsymbol{\beta}}^*$ are asymptotically normal estimates of a common $\boldsymbol{\beta}$, so $T^2$ converges in distribution to $\chi^2_q$. 
In particular $E_{H_0}[T^2] = q$, and this null distribution does not depend on the sample sizes. 
The constant $c$ is chosen so that $a = a_0$ exactly when $T^2$ equals its null expectation, i.e. $a = a_0^{T^2/q}$.

The calibration also preserves the target weight on average under the null hypothesis.
Since $E[e^{-cT^2}]$ is the $\chi^2_q$ moment generating function evaluated at $-c$
\begin{align*}
      E_{H_0}[a] &= \left(1 + 2c\right)^{-q/2}\\
                 &= \left(1 + \frac{2|\log a_0|}{q}\right)^{-q/2}
                 \longrightarrow a_0 \quad \text{as } q \to \infty,
\end{align*}
and the convergence is fast.
With $a_0 = 0.9$ and $q = 9$ as in our simulation, $E_{H_0}[a] = 0.901$. Moreover, $a$ concentrates around $a_0$ much more strongly under the $p$-value link. 
Since $T^2/q$ has null standard deviation $\sqrt{2/q}$, a first-order expansion gives 
\[\mathrm{sd}_{H_0}(a) \approx a_0 |\log a_0| \sqrt{2/q} \approx 0.045.\] 
By contrast, the $p$-value link yields a null weight with mean $0.5$ and standard deviation $1/\sqrt{12} \approx 0.29$ regardless of sample size.

When $\boldsymbol{\delta} = \boldsymbol{\beta} - \boldsymbol{\beta}^* \neq
\boldsymbol{0}$, the $T^2$ statistic is asymptotically noncentral $\chi^2_q(\lambda)$ with noncentrality
\begin{equation*}
    \lambda = \boldsymbol{\delta}^\top
    \left[\mathrm{Cov}(\hat{\boldsymbol{\beta}})
    + \mathrm{Cov}(\hat{\boldsymbol{\beta}}^*)\right]^{-1} \boldsymbol{\delta}.
\end{equation*}
Using the noncentral $\chi^2$ moment generating function,
\begin{equation*}
    E[a] = \left(1 + 2c\right)^{-q/2}
           \exp\!\left(-\frac{c\,\lambda}{1 + 2c}\right),
\end{equation*}
so the expected weight decays exponentially in $\lambda$ at rate $c/(1+2c) \approx c$ when $c$ is small. 
Because the covariance matrices shrink at rate $1/n$, $\lambda$ grows linearly in the sample sizes for any fixed $\boldsymbol{\delta}$.
Hence $a \to 0$ as evidence of incompatibility accumulates. 
The decay is controlled, however, and $\log a$ decreases by $c = |\log a_0|/q \approx 0.012$ per unit of $T^2$, whereas the $p$-value decreases by approximately $1/2$ per unit of
$T^2$ (up to polynomial factors, from the $\chi^2$ tail), a rate $q / (2|\log a_0|) \approx 43$ times faster. 
The exponential link therefore preserves the qualitative behavior of the $p$-value rule by leveraging the NPS under compatibility and discarding the NPS under strong bias. 

The properties above are asymptotic and approximate in several respects and two distinct limits are involved. 
The $\chi^2_q$ and noncentral $\chi^2_q$ distributions hold as the two sample sizes grow with $q$ fixed and assume the working logistic models are correctly specified. 
Given these distributions, the expressions for $E_{H_0}[a]$ and $E[a]$ are exact for every $q$, but the concentration of $a$ around $a_0$ improves only as $q$ grows, and the stated standard deviation is a first-order approximation.

The mean calibration is insensitive to $q$.
The expression for $E_{H_0}[a]$ gives $0.909$ even at $q = 1$. 
However, the relative null dispersion
of $T^2/q$ scales as $\sqrt{2/q}$, so for very small $q$ the null distribution of $a$ is right-skewed and noticeably dispersed. 
At $q = 9$, $80\%$ of the null mass falls in $[0.84, 0.95]$, whereas at $q = 1$ the exact null standard deviation is $0.11$ and the corresponding interval widens to $[0.75, 1.00]$. 
The weight never behaves as poorly as the uniform $p$-value, but the tight concentration reported above should not be expected when the working model has only one or two coefficients. 
Relatedly, the decay-rate advantage over the $p$-value, $q/(2|\log a_0|)$, itself scales with $q$ (about $4.7$ at $q = 1$ versus $43$ at $q = 9$), so the benefits of the exponential link are most pronounced for working models of moderate dimension as in our simulation study.

Finally, the functional form $\exp(-cT^2)$ is a pragmatic choice anchored at the null mean of $T^2$. 
A fully Bayesian treatment of $a$ is a natural refinement, which we leave to future work.

\section{Comparison of IPW estimators}\label{app:ipw}

\begin{figure*}[th]
\centering
\includegraphics[width=.9\textwidth]{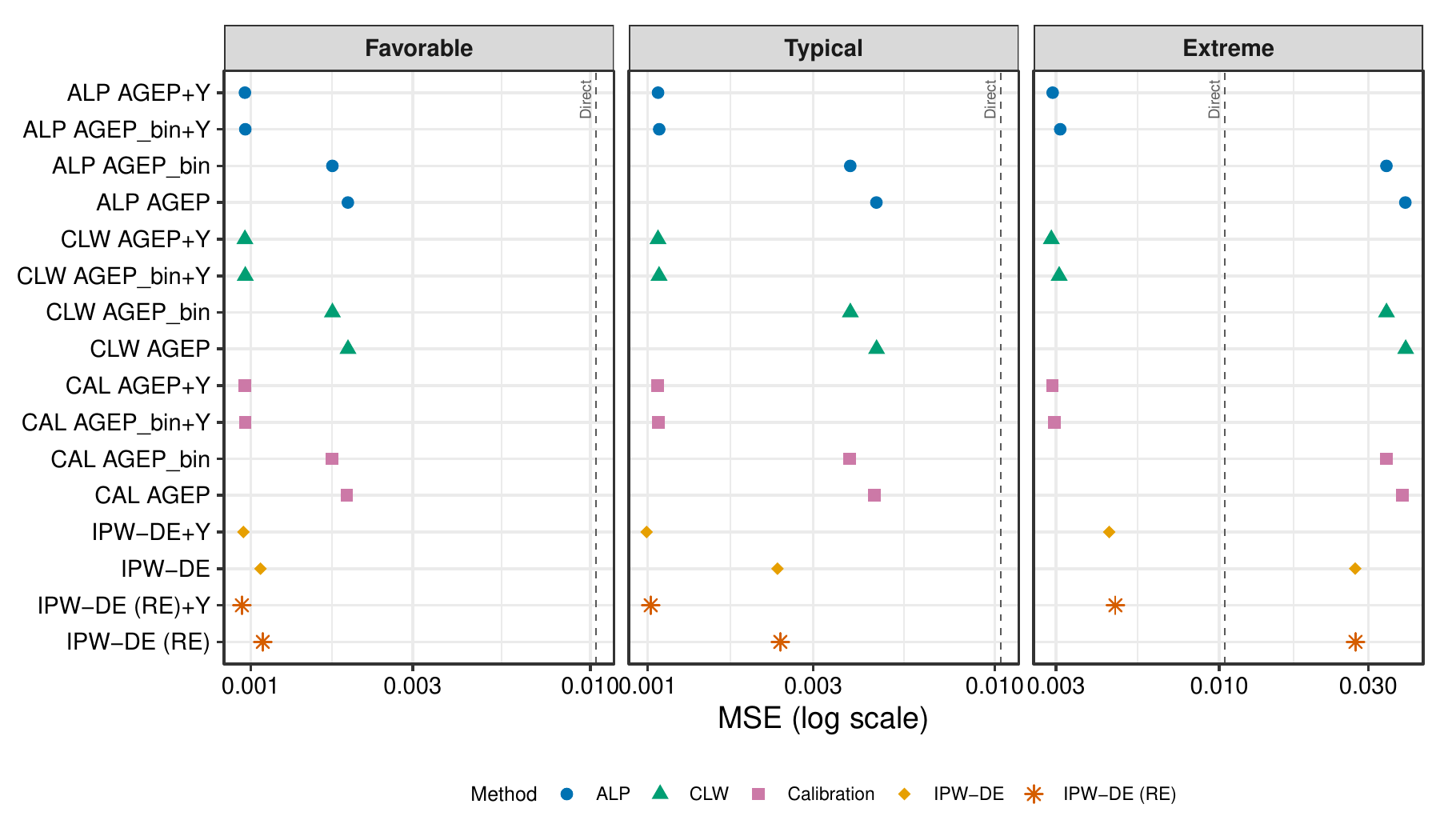}
\caption{MSE on log scale for each IPW method faceted by DDC setting.
Dashed vertical lines indicate PS-only benchmarks.}
\label{fig:app_ipw_mse}
\end{figure*}

\begin{figure*}[th]
\centering
\includegraphics[width=.9\textwidth]{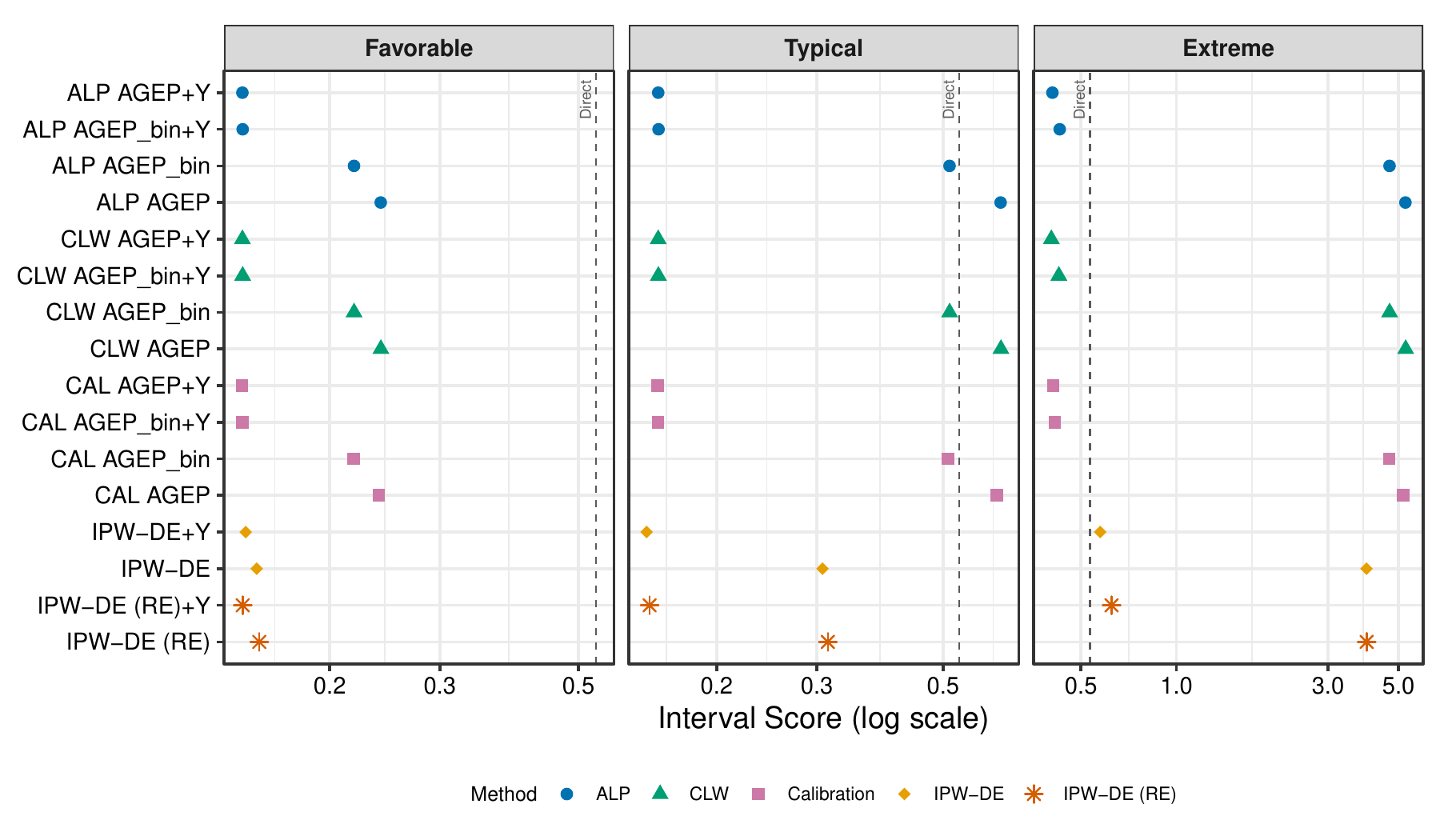}
\caption{
Interval score on log scale for each IPW method, faceted by DDC setting.
Lower values indicate better performance.
Dashed vertical lines indicate PS-only benchmarks.
}
\label{fig:app_ipw_is}
\end{figure*}

\begin{figure*}[th]
\centering
\includegraphics[width=.9\textwidth]{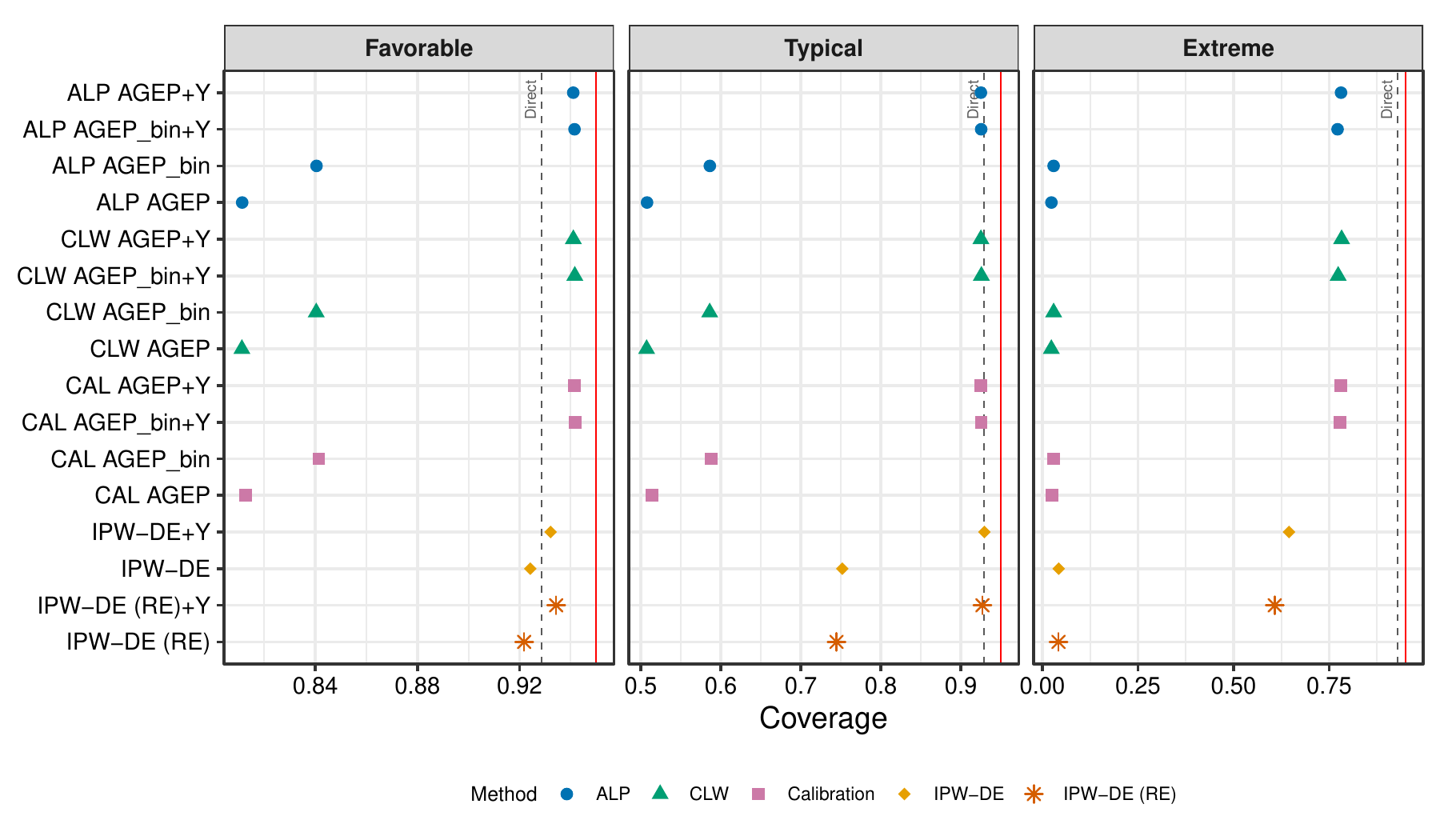}
\caption{Mean interval coverage for each IPW method faceted by DDC setting.
Nominal coverage is 95\%, marked by a solid line.
Dashed vertical lines indicate PS-only benchmarks.}
\label{fig:app_ipw_coverage}
\end{figure*}

This appendix compares the various IPW estimators summarized in Section~\ref{sec:ipw}. 
We compare direct estimates calculated using the ALP \citep{wang_2021}, CLW \citep{chen_et_al2019}, and CAL \citep{landsman_2026} methods, using the \texttt{nonprobsampling} library \citep{nonprobsampling}.
Within each method, we consider age on its original scale (AGEP) versus binned into four categories as in the main simulation (AGEP\_bin) as well as using PUBCOV as a covariate (+Y) or not.
We also include IPW-DE and IPW-DE+Y as considered in the main text along with these same models but with a random effect for PUMA.
Figures~\ref{fig:app_ipw_mse}-\ref{fig:app_ipw_coverage} display MSE, IS, and CI coverage averaged over 100 simulations (same samples as main simulation).
The choice of IPW estimator has minimal effect and all estimators with the same covariate specification are nearly identical across all three metrics.


\section{Additional Results}\label{app:results}
This appendix contains the full numerical results of the simulation study as Tables~\ref{tab:mse} and \ref{tab:interval}.

\begin{table}[!h]
\centering
\caption{\label{tab:mse}
MSE ($\times 10^3$) by DDC setting.
Results from 100 replications.
Bold indicates lowest MSE within each setting.}
\centering
\fontsize{9}{11}\selectfont
\begin{tabular}[t]{llccc}
\toprule
\multicolumn{2}{c}{ } & \multicolumn{3}{c}{DDC Setting} \\
 & Method & Favorable & Typical & Extreme\\
\cmidrule(l{3pt}r{3pt}){3-5}
\midrule
 & Direct & 18.1 & 18.1 & 18.1\\
 & BULM-PS & 3.2 & 3.2 & 3.2\\
\midrule
 & IPW-DE & 1.6 & 3.5 & 30.6\\
 & IPW-DE+Y & 0.9 & 1.0 & 2.8\\
 & IPW-DE-CLIP & 1.7 & 3.7 & 33.6\\
 & IPW-BULM & 1.1 & 2.9 & 28.3\\
 & IPW-BULM+Y & \textbf{0.5} & \textbf{0.6} & \textbf{1.6}\\
\midrule
 & MRP-INT-P-IPW & 1.2 & 3.2 & 31.4\\
 & MRP-INT-P & 1.2 & 3.0 & 30.6\\
\midrule
 & NIP (exp link) & 1.1 & 2.5 & 3.2\\
 & NIP (p-value) & 3.1 & 3.2 & 3.2\\
\midrule
 & VSW & 5.5 & 6.4 & 11.7\\
\bottomrule
\end{tabular}
\end{table}
\begin{table}[!h]
\centering
\caption{\label{tab:interval}
Coverage (Interval Score) by DDC setting.
Results from 100 replications.
Nominal coverage is 95\%.
Bold indicates lowest IS within each setting.}
\centering
\fontsize{9}{11}\selectfont
\begin{tabular}[t]{llccc}
\toprule
\multicolumn{2}{c}{ } & \multicolumn{3}{c}{DDC Setting} \\
 & Method & Favorable & Typical & Extreme\\
\cmidrule(l{3pt}r{3pt}){3-5}
\midrule
 & Direct & .905 (.826) & .905 (.826) & .905 (.826)\\
 & BULM-PS & .950 (.255) & .950 (.255) & .950 (.255)\\
\midrule
 & IPW-DE & .852 (.205) & .612 (.466) & .032 (4.331)\\
 & IPW-DE+Y & .939 (.144) & .926 (.154) & .790 (.388)\\
 & IPW-DE-CLIP & .855 (.208) & .606 (.485) & .030 (4.587)\\
 & IPW-BULM & .808 (.174) & .437 (.509) & .001 (4.593)\\
 & IPW-BULM+Y & .949 (\textbf{.105}) & .936 (\textbf{.113}) & .770 (.291)\\
\midrule
 & MRP-INT-P-IPW & .782 (.191) & .382 (.581) & .000 (4.968)\\
 & MRP-INT-P & .771 (.192) & .322 (.629) & .000 (5.000)\\
\midrule
 & NIP (exp link) & .902 (.143) & .797 (.238) & .948 (\textbf{.254})\\
 & NIP (p-value) & .946 (.254) & .949 (.255) & .949 (.255)\\
\midrule
 & VSW & --- & --- & ---\\
\bottomrule
\end{tabular}
\end{table}
\clearpage

\bibliographystyle{apalike}
\bibliography{bibliography}    

\end{document}